\documentclass[ twocolumn, 
nofootinbib,
 amsmath,amssymb,
 aps, physrev,
]{revtex4-2}

\usepackage{graphicx}
\usepackage{dcolumn}
\usepackage{bm}
\usepackage{hyperref}

\usepackage{xcolor}
\usepackage{braket}

\DeclareMathOperator*{\argmax}{arg\,max}

\begin{document}

\preprint{APS/123-QED}

\title{\textbf{A General Solution for Network Models with Pairwise Edge Coupling} 
}%

\author{Alessio Catanzaro}

\affiliation{
 Networks Unit, IMT School for Advanced Studies Lucca, Italy
}%
\affiliation{
 Lorentz Institute for Theoretical Physics, Leiden University, the Netherlands
}%
\affiliation{
 Università di Palermo, Italy
}%

\author{Subodh P. Patil}

\affiliation{
 Lorentz Institute for Theoretical Physics, Leiden University, the Netherlands
}%

\author{Diego Garlaschelli}

\affiliation{
 Networks Unit, IMT School for Advanced Studies Lucca, Italy
}%
\affiliation{
 Lorentz Institute for Theoretical Physics, Leiden University, the Netherlands
}%

\date{\today}

\begin{abstract}
Network Models with couplings between link pairs are the simplest models for a class of networks with Higher Order interactions. In this paper we give an analytic, general solution to this family of Random Graph Models extending previous results obtained for specific interaction structures. We use the Hubbard-Stratonovich transform to show that such higher order models can be decoupled into families of effectively separable random graph models, which might be enhanced with auxiliary, possibly multi-valued, parameters  identifying the thermodynamical phases of the system. We study the diagonalizability of the matrix of the interactions between links, that is itself an adjacency matrix on the dual network whose vertices are the nodes pairs, that can be dealt with using tools of algebraic graph theory. By orthogonalizing the interaction,  we carry out the full computation of the partition functions and discuss why in some cases interactions effectively reduce to their Mean-Field versions.
\end{abstract}

\maketitle

\section{Introduction}
Complex system science is heavily reliant on networks to model a multitude of systems. Random networks are among the most used tools to encode both the random nature of such objects as well as their evident heterogeneity. A statistical mechanics approach to random networks has been widely used by the network science community in the last 20 years \cite{park2004statistical, cimini2019statistical, squartini2017maximum, coolen2017generating}. Yet a fundamental hurdle in the modelling of such systems with the aforementioned approaches is that few among the innumerable conceivable models of random networks can be solved, and if so it is often through approximate techniques that are model-dependent.

First order networks are easily solved, as the Hamiltonian describing them has only link specific parameters, which gives rise to a separable partition function and hence independent, link-specific probabilities. Regretfully when any form of coupling among links are introduced one has often to resort to either trivial toy-models of the interaction, or to be satisfied with approximated solutions only\cite{park2004statistical, park2004solution}.

The purpose of this paper is to partially cover this knowledge gap by showing how a general solution for any second order unweighted, undirected random network can be written, and to solve for some cases of interest by showing some non-trivial behaviour. We also note that the analytical techniques we use could be extended to provide for a solution of weighted and directed random networks, but we do not believe the physical behaviour of such models to be fundamentally different than the one we present here in terms of phase transitions.

The motivation for treating pairwise-coupled link models as a tractable
limit of higher-order network theory is strengthened by the
shift the network science community has taken in the last decade,
moving from a purely dyadic description of complex systems towards
representations that explicitly account for group interactions, such as
hypergraphs and simplicial complexes~\cite{Battiston2020,Boccaletti2023}.
Within that broader framework, pairwise coupling between edges constitutes the simplest nontrivial departure from first order network models, and can be regarded as a controlled
approximation to genuinely triadic or higher interactions whenever the
correlation between two edges can be meaningfully attributed to a single
shared structural cause (a common node, a shared community), as opposed to an irreducible collective effect among three or more edges. In this sense, the class of models we solve
occupies a precise and useful niche: complex enough to generate genuine
phase structure, yet simple enough to remain solvable, and
therefore well suited to serve as a null model against which genuinely
higher-order effects can later be isolated (see also section \ref{sec:meanfield}).

Our approach relies on the implementation of the multidimensional Hubbard-Stratonovich transform to account for any coupling structure between links. This also implies an effective description of the model through independent links, conditional on a given order parameter that can be treated as a random variable (and being a function of the model given parameters). Our solution pipeline also extends some results already found in the literature in order to put them under a single umbrella of explicative framework. 

A further motivation for obtaining such a general solution comes from network inference. Ordinarily, exponential random graph models with pairwise dependence between links have an intractable normalizing constant, which forces practitioners to resort to approximate estimationschemes such as Markov Chain Monte Carlo maximum likelihood \cite{Snijders2002} or related curved exponential-family techniques \cite{HunterHandcock2006}, both of which are known to be computationally expensive and prone to model degeneracy \cite{Handcock2003}. Because the solution we derive gives explicit access to the partition function and hence to the full likelihood of any second order model in this class, it removes this bottleneck: the log-likelihood of an observed network can be evaluated, and in many cases maximized, directly and exactly, rather than estimated through simulation. This places second order, pairwise-coupled models on the same operationally tractable footing that first order maximum-entropy models already enjoy for network reconstruction and null-model based inference \cite{squartini2017maximum}.

The paper is organized as follows: in Section \ref{sec:GenSol} we introduce the techniques to solve the general problem, the pipelines to compute the partition function and write the single link probability function in an exact, although not explicit, manner. In Section \ref{sec:examples} we show how the machinery works for some cases of second order Random Network Models (RNM) to give meaning to our formulae, and we give explicit solutions for a few models of interest. Some models for which we solve were actually already found in the literature, but as stand-alone works and without the general, explicative framework we provide. Moreover, in the case of the model of \ref{sec:twostar}, we highlight a connection between the model already analyzed in \cite{annibale2015two, park2004solution} to the field of Coding Theory \cite{delsarte1973algebraic}. Finally in Section \ref{sec:meanfield} we comment on how to give a rule of thumb to determine the possible phase structure of the models at stake and if the mean field approximation can give a satifying picture of the whole phase space.

All the derivations of the formulae we show are given in a nutshell in their respective Section, while the thorough computations are moved to the Supplementary Information (SI) file for the sake of brevity of the paper.

\section{Solution to II order Random Network}

\label{sec:GenSol}

A Random Network Model (RNM) is a probability law $P(\mathbf{A})$ for the realizations of the adjacency matrix $\mathbf{A}$ of a network. Each entry $A_{ij}$ of the matrix describes if and how nodes $i$ and $j$ are connected, with (conventionally) $0$ representing the absence of a connection, and a different value to indicate the presence of an interaction. We choose to deal with unweighted, undirected networks so that $\mathbf{A}$ is a symmetric matrix and its entries $A_{ij}$ can be either $0$ or $1$. This latter fact implies the fundamental relation

\begin{equation}
    A_{ij} = A_{ij}^m \quad  m  = 1, 2, 3, \ldots \label{eq:commutA}
\end{equation}

that simplifies enormously the computations that follow along the entire paper as we show in SI. 
For instance, we can use Relation \eqref{eq:commutA} to show that we can represent the probability function describing the model in the exponential form 

\begin{equation}
    P(\mathbf{A}) = \frac{e^{-H(\mathbf{A})}}{Z} \qquad Z = \sum_{\{\mathbf{A}\}}e^{-H(\mathbf{A})} \label{eq:ERG}
\end{equation}

where $\{ \mathbf{A} \}$ indicates all the possible configurations of the matrix $\mathbf{A}$. The Hamiltonian is in the form to eventually enforce all the possible (up to $p$) constraints

\begin{equation}
    H(\mathbf{A}) = -\sum_{l=1}^{p} \theta_{l} f_l(\mathbf{A})\label{eq:3}.
\end{equation}
Here, \(f_l(A)\) denotes the \(l\)-th network property whose ensemble average is constrained, such as the number of links, node degrees, or counts of specified subgraphs. The parameter \(\theta_l\) is the Lagrange multiplier conjugate to \(f_l\), controlling its expected value in the ensemble.

If one desires to enforce the full ERG formalism \cite{park2004statistical, cimini2019statistical, squartini2017maximum, coolen2017generating}, then one has to add the constraint equations such that

\begin{equation}
    \sum_{\{\mathbf{A}\}} f_{l}(\mathbf{A})P(\mathbf{A}) =  f_{l}^*\quad\forall l,
\end{equation}
where $f_{l}^*$ is the `target' expected value of the property $f_l$, which can be interpreted in two ways. The first interpretation is to take $f_{l}^*$ as the empirical value $f_l^* = f_l(\mathbf{A}^*)$ from a real observed matrix $\mathbf{A}^*$.
A second interpretation is  one where the parameter space of the model is explored exhaustively and $f_l^*$ is the expected value of the property over the ensemble implied by the chosen parameters. We will follow this latter interpretation throughout this work, but we also notice that this work constitutes a step towards the use of II order RNM as null models for inference on random networks. Indeed, the explicit computation of the Partition function we will be able to perform allows the usage of maximum likelihood estimators on graph instances.
 {Using the binary identity in Eq. \eqref{eq:commutA}, the Hamiltonian can be expanded as a multilinear function of the adjacency-matrix elements}
\begin{eqnarray}
    H(\mathbf{A}) =  &-& h^{(0)} - \theta_1\sum_{ij} h^{(1)}_{ij} A_{ij} - \theta_2 \sum_{i,j,k,l} h^{(2)}_{ijkl} A_{ij}A_{kl} \label{eq:multilinearexpression} \nonumber \\
    &-& \theta_3\sum_{i,j,k,l,m,n } h^{(3)}_{ijklmn} A_{ij}A_{kl}A_{mn} - \ldots 
\end{eqnarray}
 {where we denote the coefficients of the terms of different interaction orders by \(h^{(0)},h^{(1)},h^{(2)},\ldots\), and reabsorb different Lagrange multipliers tied to observables of the same order in a single coefficient $\theta_l$ which is encompassing all the network features of that order.} The above expansion is analogous to similar computations performed in the context of QFT for Grassmannian variables \cite{rogers2007supermanifolds}, where the Relation \eqref{eq:commutA} is replaced by the anticommuting relation. We leave the detailed proof of these claims to Supplementary Information (SI) for the sake of brevity.

In the non-interacting case, where $\theta_2 = \theta_3 = \ldots = 0$ in the Hamiltonian \eqref{eq:multilinearexpression}, the probability for the single link is easily solved as a result of the separation of the overall probability, which reduces to the Bernoulli form 

\begin{eqnarray}
    P(\mathbf{A}) &=& \frac{e^{h^{(0)}}}{Z} \prod_{i<j}e^{\theta_1 h^{(1)}_{ij} A_{ij}} \nonumber \\
    &=& \prod_{i< j} \left(\frac{p_{ij}}{1-p_{ij}}\right)^{A_{ij}}(1-p_{ij}) \nonumber \\
    &=& \prod_{i< j} p_{ij}^{A_{ij}} (1-p_{ij})^{1-A_{ij}}  
\end{eqnarray}

where $p_{ij} = \frac{e^{\theta_1 h^{(1)}_{ij}}}{1 + e^{\theta_1 h^{(1)}_{ij}}}$ and $\frac{e^{h^{(0)}}}{Z} = \prod_{i<j}\left(1-p_{ij}\right)^{-1}$. In this case $Z$ as well is separable and becomes

\begin{equation}
    Z = \prod_{i<j}z_{ij} = \prod_{i<j}\sum_{A_{ij}=0,1} e^{\theta_1 h^{(1)}_{ij} A_{ij}}.
\end{equation}

According to the specific functional form of $h^{(1)}_{ij}$ one can retrieve several specific models. The most notable include Erd\"os-Renyi, when $h^{(1)}_{ij} = h^{(1)}$; the Configuration Model, in which $h^{(1)}_{ij} = h^{(1)}_{i} + h^{(1)}_{j}$; and Stochastic Block Model, when $h^{(1)}_{ij} $ assumes the right block structure so that we can rewrite the Hamiltonian as $H(A) = -\sum_{p,q}\epsilon_{pq}\sum_{i\in p, j\in q} A_{ij}$ where $p$ and $q$ are the labels of the different communities the network is partitioned into.

The purpose of this paper is to show how a solution to the second order approximation of a model defined by the Hamiltonian \eqref{eq:multilinearexpression} can be found in general.

\subsection{Evaluation of the partition function}

Our main task is to compute the partition function, or equivalently the single link probability function of the model at stake. For the sake of computations, we first vectorize the linear part of the Hamiltonian  encoding for the local fields. This is done by applying a lexicographic ordering to the couples of node indices $(i,j)\equiv a$ so that instead of having an adjacency matrix for both the link  variable and the local parameter, we end up with a list of links and their respective parameters, going from 1 to $V=\frac{N(N-1)}{2}$.  Through this reordering, one can also express the tensor that is involved in the quadratic form of Equation \eqref{eq:multilinearexpression} as a matrix.  The reader can find the thorough derivation of this in SI, and they will be easily convinced that a compact way to represent this Harmonic Oscillator of RNMs is with the  bra-ket Hamiltonian

\begin{equation}
    H(\mathbf{A}) = -\braket{\mathbf{A}|\epsilon} - \frac{1}{2V}\braket{\mathbf{A}|\mathbf{\beta}|\mathbf{A}}\label{eq:ham}
\end{equation}

where $\mathbf{\beta}$ is the interaction matrix in which all the  couplings (and their intensity) between links is encoded, namely

\begin{equation}
    \beta_{ab} = \theta_2 h^{(2)}_{ab}.
\end{equation}
Note also that the  new, local parameter is  redefined to be $\epsilon_{ij} =\theta_1 h^{(1)}_{ij}+\frac{\beta_{ijij}}{2V}$, clearly reducing to $h^{(1)}_{ij}$ in the large $N$ limit. 

This matrix $\mathbf{\beta}$ is symmetric by construction, as it is the coupling matrix between the links,  representing a 'dual' undirected network over the couples of nodes. Hence, it can be diagonalized via the spectral theorem with a basis of orthonormal eigenvectors

\begin{equation}
    \mathbf{\beta} = \sum_{k=1}^r \lambda_{k}\ket{\omega_k}\bra{\omega_k}\label{eq:diagonalization}
\end{equation}

where $r$ is the rank of the interaction.
The Hamiltonian then can be decoupled in the sum over the eiegenmodes of the interaction and exponentiated  to compute the sum over the possible configurations. Then it is in a form by which all the squares arising from the product of the eigenvectors with each configuration can be decoupled with the Hubbard-Stratonovich transform.  This kind of manipulation has a long history in theoretical physics \cite{Hubbard1959, Stratonovich1957} and is a standard tool to decouple interacting fields through an auxiliary variable which can then be integrated over.  {The last step is to evaluate the exponential over the configurations which are now effectively decoupled, so to get the Partition function of the system}
 {
\begin{equation}
    Z = 
    \int  d\mathbf{x}^r\,
     \frac{e^{- \sum_{k=1}^r\frac{x_k^2}{2}}}{(2\pi)^{-r/2}}
    \prod_{a=1}^{V}
    \left[
    1+
    e^{\epsilon_a+
    \sum_{k=1}^{r}
    x_k\sqrt{\frac{\lambda_k}{V}}(\omega_k)_a}
    \right]\label{eq:zprod}
\end{equation}}
 {or, more compactly,}

 {
\begin{equation}
    Z = 
    \int  d\mathbf{x}^r \frac{e^{\Phi(\mathbf{x})}}{(2\pi)^{-r/2}}
\end{equation}}
 {where it is convenient to define the action term}
 {
\begin{equation}
    \Phi(\mathbf x)
=-\sum_{k=1}^{r}\frac{x_k^2}{2} +\sum_{a=1}^{V}
\ln\left[
1+e^{
\epsilon_a+\sum_{k=1}^{r}
x_k\sqrt{\frac{\lambda_k}{V}}(\omega_k)_a
}\right]
\end{equation}}
 {The appearance of the $x_k$ variables is the key ingredient of the Hubbard-Stratonovich transform, and we will see how they allow us to approximate the Partition Function in an effectively separable form, conditionally on the values that the auxiliary variable take. These last equations are one of the central results of this paper, as will be employed in the actual computations of the models in the following sections. The full derivation can be again found in SI for the sake of brevity and readability.}

\subsection{Effective Separability of the partition function}
We know from general Statistical Mechanics theory that the model defined by Equation \eqref{eq:ham} might undergo phase transitions, with a phase structure dictated by the  structure of the interaction matrix $\mathbf{\beta}$. In the Hubbard-Stratonovich approach  we follow, the multiplicity of the possible phases of the model can be encoded in the vector of values  {$\mathbf{x^*}=(x_1^*, x_2^*, \ldots x^*_{r})$},  $r$ being the maximal rank of the interaction, where each $x_k^*$ can be either $0$ or multi-valued depending on the parameters of the model.
Conditionally on the values of $(x_1^*, x_2^*, \ldots )$, the model can be rewritten in an  {effectively non-interacting form where the full partition function of Equation \eqref{eq:zprod} can be approximately factorized by retaining the first term of the effective action} (see SI) as
 {\begin{equation}
     Z \simeq \sum_{\mathbf{x^*}}\prod_{i<j}z^{\text{eff}}_{ij}(x_1^*, x_2^*, \ldots ) \label{eq:sepZ}
 \end{equation}}
 {where the single link contribution can be isolated as}
 {
\begin{equation}
        z^{\text{eff}}_{ij}(\mathbf{x^*} ) = c(\mathbf{x^*})\left[1+ e^{\epsilon_{ij} + \sum_{k=1}^r x^*_{k}\sqrt{\frac{\lambda_k}{V}}\left(\omega_k\right)_{ij}}\right]\label{eq:zsinglelink}
\end{equation}}
 {and $c(\mathbf{x^*})$ is a prefactor depending only on the value of $\mathbf{x^*}$.} 
The last expression is equivalent to that corresponding to a  modified, non-interacting ($\mathbf{\beta}=0$) version of the Hamiltonian in Equation \eqref{eq:ham}, i.e.
\begin{equation}
    H(A|x_1^*, x_2^*, \ldots) = - \braket{A|\epsilon'(x_1^*, x_2^*, \ldots)}  
\end{equation}
where we have introduced the modified parameters \begin{equation}
    \ket{\epsilon' (x_1^*, x_2^*, \ldots)} = \ket{\epsilon} + \sum_{k=1}^r x_k^* \sqrt{\frac{\lambda_k}{V}}\ket{\omega_k}.
\end{equation}

From this, it is easy to write the single link probability in terms of the model parameters, complemented with the new one coming out of the decoupling via the HS transform.

\begin{equation}
    p^{\text{eff}}_{ij} = \frac{e^{\epsilon_{ij}+ \sum_{k=1}^r x_k^* \sqrt{\frac{\lambda_k}{V}}\left(\omega_k\right)_{ij}}}{1+ e^{\epsilon_{ij}+ \sum_{k=1}^r x_k^* \sqrt{\frac{\lambda_k}{V}}\left(\omega_k\right)_{ij}}}\label{eq:probsinglelink}
\end{equation}

for the link probability expression. Formulae \eqref{eq:zsinglelink} and \eqref{eq:probsinglelink}, together with Equation \eqref{eq:zprod} are the core results of this paper.  {Before moving on, it is worthwhile to discuss the fact that the separability result of the partition function comes from a first order approximation around the peak of the exponential using Laplace technique, and in SI we also derive the correction to the second order, in which a prefactor containing the square root of the determinant of the Hessian comes in place. However, by estimating the correction (see SI) to the first order approximation, one realizes that the only case in which the second order correction has to be considered is when the rank $r$ of the interaction grows with the square of the system $V \sim N^2$. We will analyze one case in which this happens in Section \ref{sec:twostar}, that becomes nonetheless solvable thanks to its regularity. In the other cases, and in most of practical instances, the first order approximation is good enough to capture the behaviour of the partition function.}

\subsection{Phase structure}
The price to pay  to write Equations \eqref{eq:zsinglelink} \eqref{eq:probsinglelink} with such a product over node couples and to have an effectively separable partition function over links is that a phase structure emerges.  In our case we encode this behaviour in the parameter  {$\mathbf{x^*}$}. In the derivation of Equation \eqref{eq:sepZ} (see SI),  {we understand that $\mathbf{x^*}$ is the point for which the gradient of the logarithm of the partition function is vanishing, identifying the critical (maxima, with the additional condition that the Hessian of the action must be negatively definite) values of the auxiliary fields that allow to perform a saddle point evaluation of eq. \eqref{eq:zprod}. Namely, this means the mathematical condition}

\begin{figure}[h!]
\includegraphics[width=\linewidth]{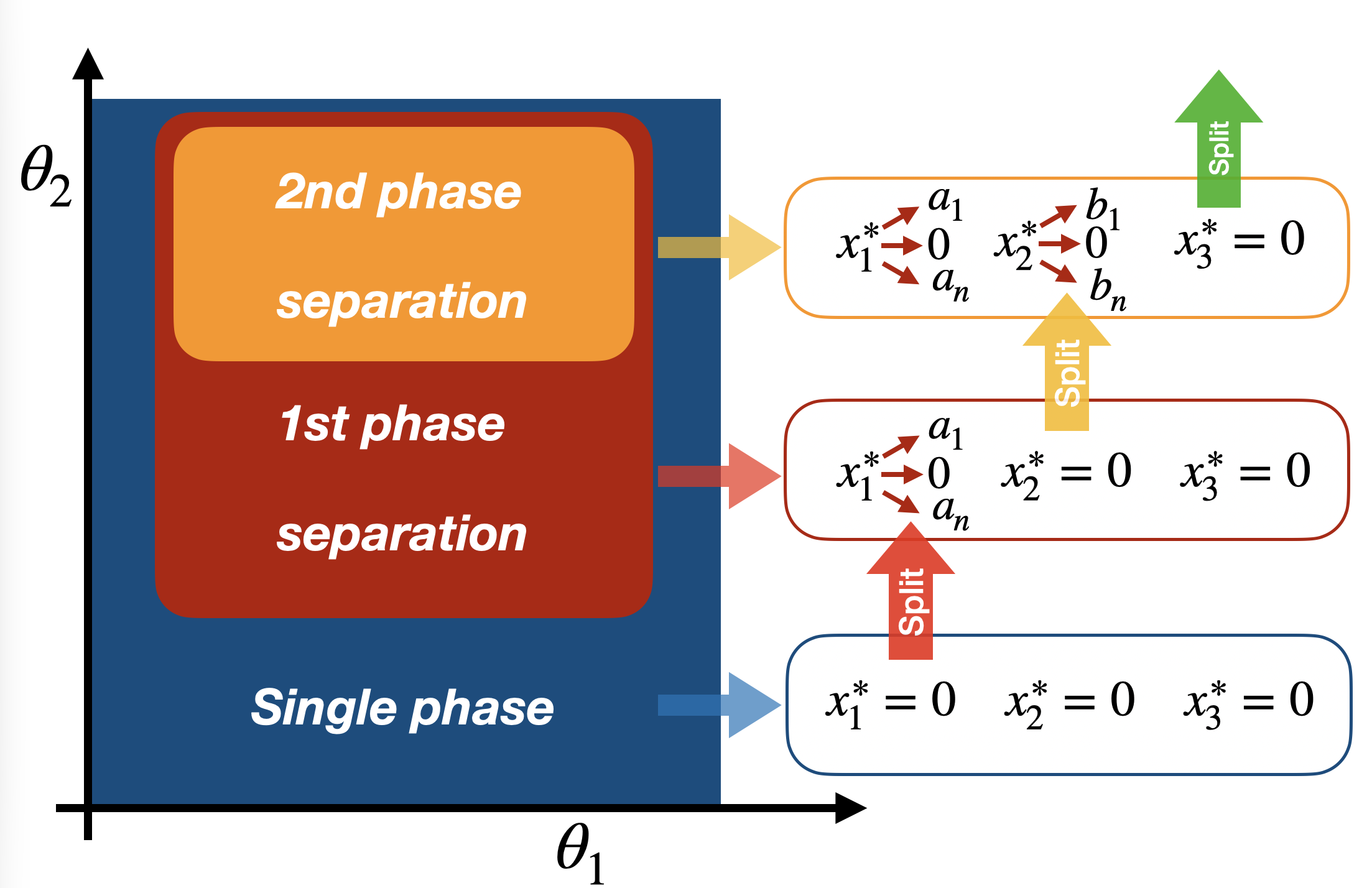}
\caption{\label{fig:phasetrans}A schematic representation of the phase diagram of a II order RNM and the parameter $x_k^*$  in the sparse case (large, negative $\epsilon_{ij}$ implying $x_{k}^*\simeq0$). One has to look if the region of the parameter space whom is looking at may be of single phase or with phase separation. This is done by looking at the multiplicity of solutions of the equation defined by  {putting the gradient of $\Phi(\mathbf{x})$} equal to zero. }
\end{figure}

 {\begin{equation}
    \frac{\partial\Phi}{\partial x_k}|_{\mathbf{x}^*} =0  \iff x_k^* = \sum_{i <j}\sqrt{\frac{\lambda_k}{V}}\left(\omega\right)_{ij}p_{ij}(\mathbf{x_*})\label{eq:stationar}
\end{equation}}
 {for all modes $k$.} This new parameter  {vector $\mathbf{x^*}$} is global and depends only on  the specific structure of the interaction, combined with the local fields $\epsilon_{ij}$, and most importantly its unicity is not guaranteed. Actually, when the  {system} parameters combination allows, $x_k^*$ splits into multiple values (found numerically) that, when insterted in Equation \eqref{eq:probsinglelink}, give rise to different values of the single link probability. A scheme of what happens is shown in Figure \ref{fig:phasetrans}.

In the cases in which the parameter splits, the actual value of the link probability changes, hence identifying different phases (possibly denser or sparser). When drawing a realization of the ensemble, one has to keep in mind that also the various values of $x^*_k$ come each with its own probability and can be computed as we show in Section \ref{sec:PhaseCoex}.

\section{II order interacting models}
\label{sec:examples}
We will now show how Equations \eqref{eq:zprod}, \eqref{eq:zsinglelink} and \eqref{eq:probsinglelink} can be used to actually solve some II order RNMs. We will start with a very simple, homogeneous one to use as a lamppost for the general behaviour and explore more complex models going on.

\subsection{RNM with Homogeneous interactions}
\label{sec:homo}
We start with the Hamiltonian in the form
\begin{equation}
    H(\mathbf{A}) = - \braket{\mathbf{A}|\epsilon} - \frac{1}{2V} \braket{\mathbf{A}|\mathbf{\beta}|\mathbf{A}}\quad \text{ with } \mathbf{\beta} = \theta \ket{1}\bra{1},
\end{equation}
so that every link interacts with any other link in the same way. The eigenvalues of $\mathbf{\beta}$ are all $0$ except for the eigenvalue $\lambda = V \theta$. The corresponding eigenvector is also easy to find as it is just $\ket{\omega}=\frac{\ket{1}}{\sqrt{V}}$. Moreover we set ourselves in the homogeneous field case $\epsilon_a = \epsilon$. The partition function then becomes easy to compute explicitly in the large $N$ limit, as in traditional calculations using the Hubbard-Stratonovich transform. From Equation \eqref{eq:zprod} we have
\begin{eqnarray}
    z_a^{\text{eff}} &=& \lim_{V\to \infty}\left[\sqrt{\frac{1}{2\pi}} \int \,dx e^{-\frac{x^2}{2}} \left[1+ e^{\epsilon + x \sqrt{\frac{\theta}{V}}  }\right]^V\right]^{\frac{1}{V}} \nonumber \\
    &=& \lim_{V\to \infty}\left[\sqrt{\frac{V}{2\pi}} \int \,dy \left[e^{-\frac{y^2}{2}} \left(1+ e^{\epsilon + y \sqrt{\theta} }\right)\right]^V\right]^{\frac{1}{V}} \nonumber \\
    &=& \max_{-\infty \leq y \leq \infty}\left\{e^{-\frac{y^2}{2}}(1+ e^{\epsilon + y \sqrt{\theta}})\right\} \\
    &=&  { e^{-\frac{(y^*)^2}{2}}(1+ e^{\epsilon + y^* \sqrt{\theta}}) }\label{eq:z_homo}
\end{eqnarray}

where the substitution $x = y \sqrt{V}$ has been made in the second row,  {and $y^* = \argmax\left\{e^{-\frac{y^2}{2}}(1+ e^{y \sqrt{\theta} + \epsilon})\right\}$}. The  {third} step comes from a Laplace saddle point evaluation  {(here in the formulation of \cite{polya1972problems})}.

\begin{figure}[h]
\includegraphics[width=\linewidth]{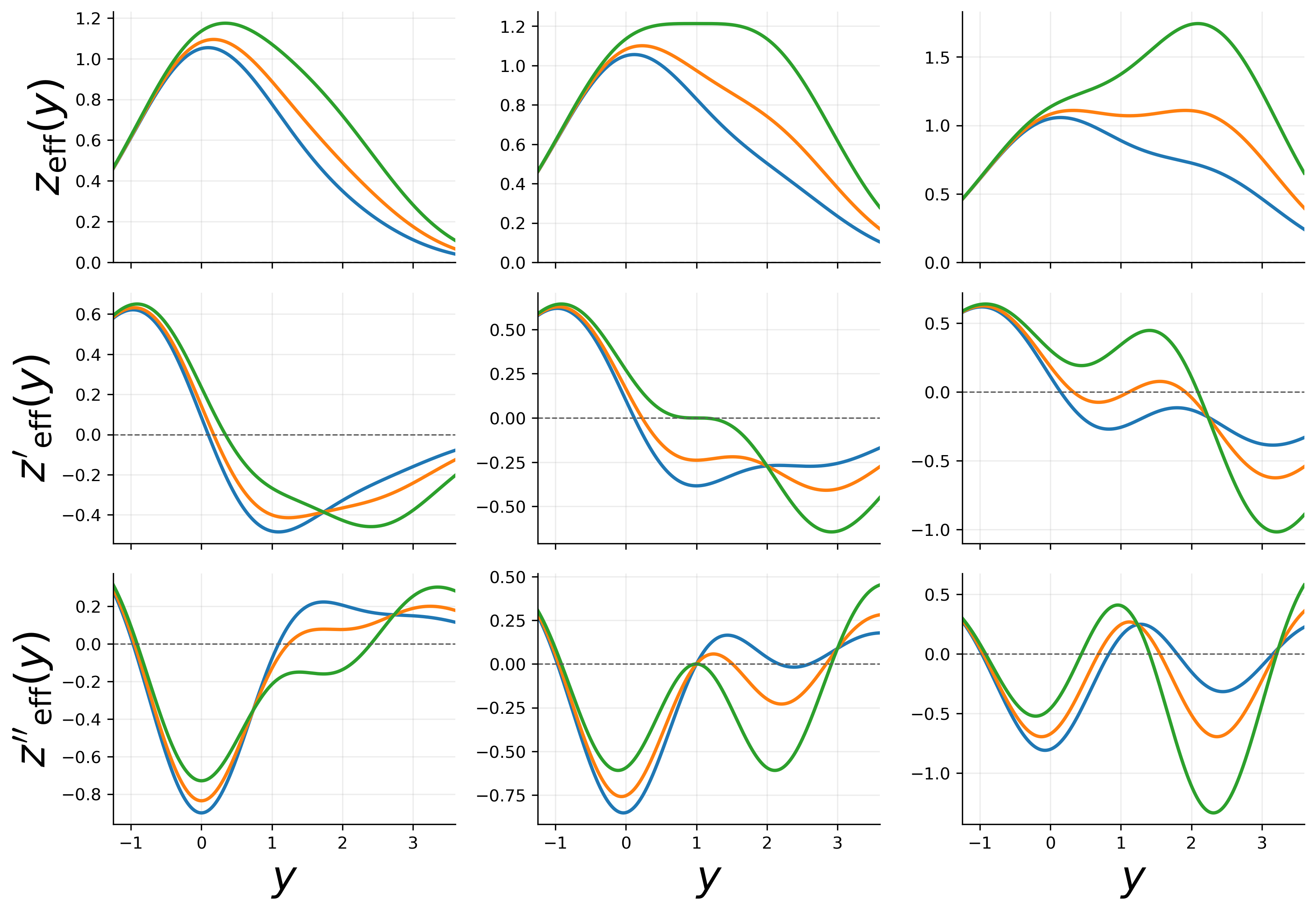}
\caption{\label{fig:graphical_solution}Study of the function $z^{\text{eff}}$ for different values of $\epsilon, \theta$. From top to bottom we plot $z^{\text{eff}}$, its first and second derivative to locate the maxima and show that there is a regime in which 2 maxima are present. The parameter $\epsilon$ always takes the  {values $-3$ (blue line), $-2.5$ (orange line), $-2$(green line),} while $\theta$ is, from left to right $3, 4, 5$ respectively. }
\end{figure}

\begin{figure}[h]
\includegraphics[width =\linewidth]{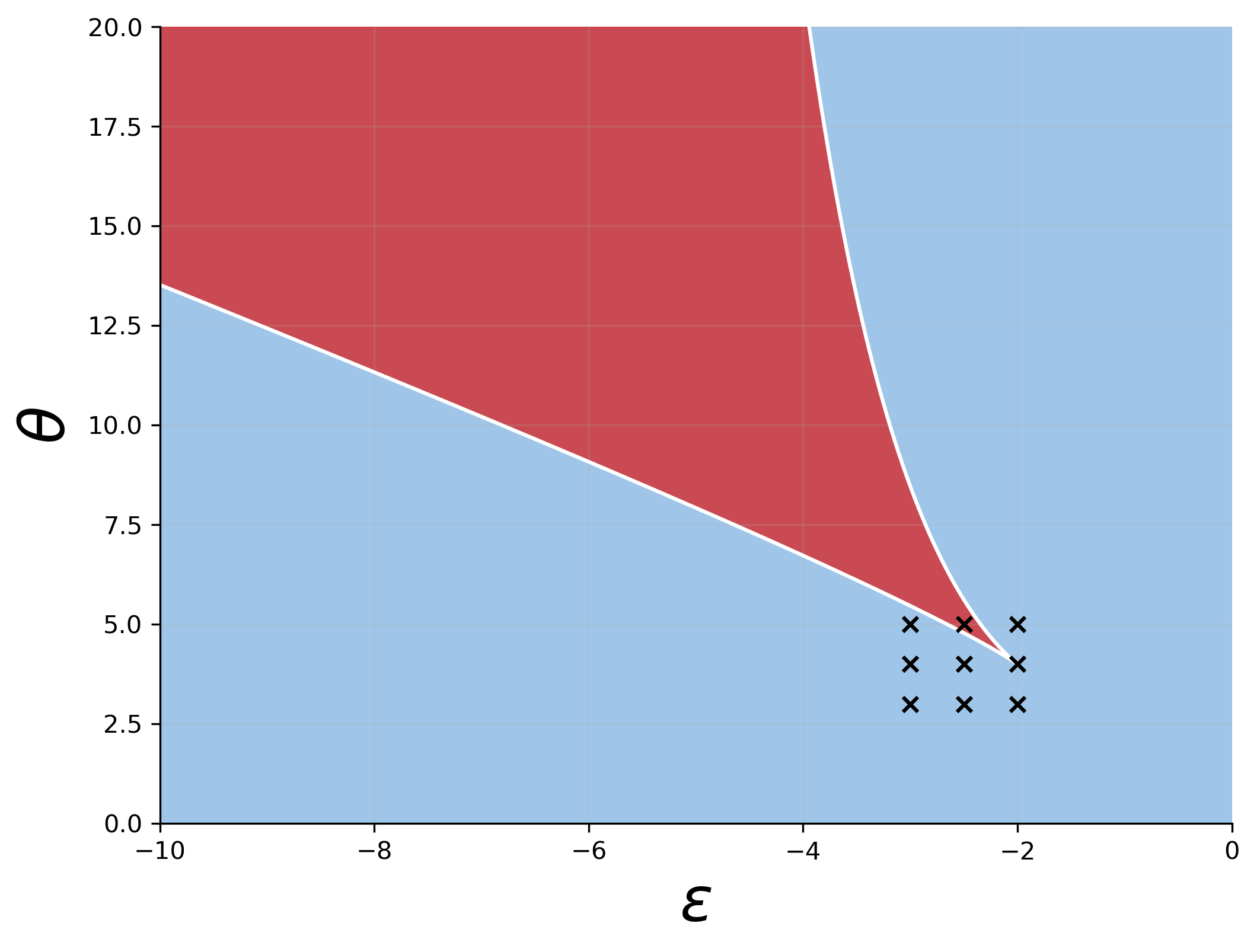}
\caption{\label{fig:PD_homo} Phase diagram for the  homogeneous system. The blue region corresponds to the place where there is only one maximum, hence only one link probability. In the red region the maxima of Equation \eqref{eq:z_homo} are two and the link probabilities as well. The crosses indicate the coordinates of Fig. \ref{fig:graphical_solution}}
\end{figure}

To evaluate the $z^{\text{eff}}$ in the large limit we need to see where its derivative vanishes and its second derivative is negative, which can be done graphically and is shown in Figure \ref{fig:graphical_solution}. We can see that for $\theta \geq 4$ there is a range of $\epsilon$ for which the partition function has more than one maximum. Indeed, in the parameter space we can locate a whole region in which the maxima are two and hence there are two possible values of the link probability. This is shown in Figure \ref{fig:PD_homo},  {which is drawn by inserting the critical value of $y^*$ in the single link probability function \eqref{eq:probsinglelink}, which in this case reduces to}

 {\begin{equation}
    p_{eff}(y^*)= \frac{e^{\epsilon + y^*\sqrt{\theta}}}{1+ e^{\epsilon + y^*\sqrt{\theta}}}
\end{equation}}

\subsubsection{Phase coexistence}
\label{sec:PhaseCoex}
\begin{figure}[h]
    \includegraphics[width=\linewidth]{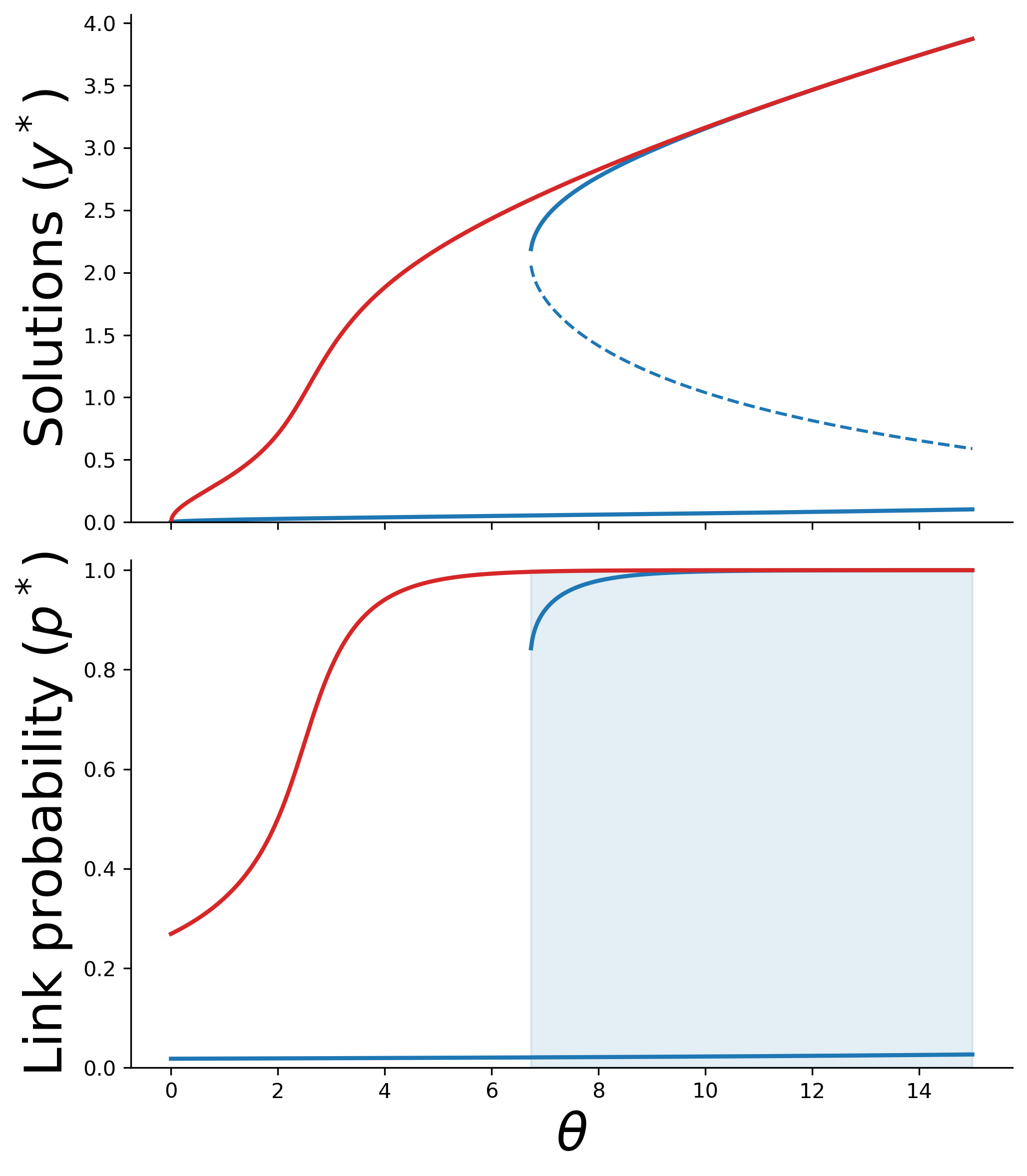} 
\caption{\label{fig:cvsb} Possible solutions of equating to zero the derivative (top panel,  dashed line indicate minima) of \ref{eq:z_homo} and effective link probability  {from \eqref{eq:probsinglelink}}(bottom panel,  shaded region is where two solutions are possible) vs $\theta$ from $0$ to $15$ with $\epsilon = -4$ (blue) and $\epsilon = -1$ (red).}
\end{figure}

In Figure \ref{fig:cvsb} we can see the effective link probability for two slices of the phase diagram, one across the single phase region (red, $\epsilon = -1$) and another one across the single phase, then a bifurcation point, then again single phase (blue, $\epsilon = -4$).
The key ingredients for the analysis of this simple, interacting model are now in place.  {If we want to write the single link effective probability as in Eq. \eqref{eq:probsinglelink}, we have to be careful to the fact that $x^*$ has multiple values.} From Figure \ref{fig:cvsb} it is clear that when $\theta\gg 4$ and $\theta < 4$ only one phase is present, the sparse one and the dense one respectively, while in the regime in which $\theta \gtrsim 4 $ the two phases coexist.

This behavior can be effectively described by the scheme in which, every time a realization of the networks has to be generated, a coin is tossed. This coin determines if the realization will be a dense or a sparse one, \textit{i.e.} this first coin toss chooses the (later) coin that will be tossed for each link in the realization. Then, when another realization has to be generated, this phase coin is tossed again and so on.  For each instance of the network there is a preliminary coin toss that determines its phase.

The relative abundance of the two phases (and hence the weights of the phase coin) can be computed in the large $V$ limit by plugging the respective values of $p$ in the formula for the overall $P(\mathbf{A})$. For instance, for the dense phase in which the probability assumes the higher value $p_d \equiv \max \left\{ p^{\text{eff}} \right\}$  we can write the probability that the overall number of links is the one we expect in the dense phase, that in the homogeneous case is simply $Vp_{d}$. As we know the probability now factorizes, we get

\begin{widetext}

\begin{equation}
    P \left\{ \sum_{i< j}A_{ij}
    = Vp_d\right\} 
    \propto \binom{V}{V\frac{1+p_d}{2}} e^{\epsilon V  p_d + \frac{\theta}{2}V p^2_d}
    \simeq \exp{ \left\{ V\left[ \epsilon p_d + \frac{\theta}{2}p^2_d +\frac{1-p_d}{2}\ln \frac{2}{1-p_d} + \frac{1+p_d}{2} \ln  \frac{2}{1+p_d}\right]\right\}} \label{eq:phaseabundance1}
\end{equation}

\end{widetext}

 where the second line follows by applying Stirling approximation. For the sparse phase, we just need to replace the $p_d$ with the lower single-link effective probability $p_s \equiv \min \left\{ p^{\text{eff}} \right\}$. This allows us to readily set the proportionality constants as we only have those two phases

\begin{equation}
    P \left\{ \sum_{i< j}A_{ij}=Vp_d\right\} + P \left\{ \sum_{i\leq j}A_{ij}=Vp_s\right\} = 1. \label{eq:phaseabundance2}
\end{equation}

Finally, we can write the original probability in the effective, non-interacting form as

\begin{eqnarray}
    P(A) &=& \prod_{i<j} p_d^{A_{ij}}(1-p_d)^{1-A_{ij}}P \left\{ \sum_{i< j}A_{ij}=Vp_d\right\} \nonumber \\
    &+& \prod_{i<j} p_s^{A_{ij}}(1-p_s)^{1-A_{ij}}P \left\{ \sum_{i< j}A_{ij}=Vp_s\right\}
\end{eqnarray}

so that we see the model can be effectively decoupled into two different, effectively independent homogeneous models, conditional on the probability that one of the two phases is realized.

Before moving to more complicated versions of II order Random Network Models, it is the right moment to discuss the relation between the models we are discussing and the Ising model. Indeed, the technique we are using to write the effective link probabilities are the same employed in the solution of the Annealed version of the Ising Model, for example in the Curie-Weiss solution where an all-to-all coupling is enforced \cite{Friedli_Velenik_2017, Kochmanski_2013}. This class of models can be mapped into an Ising Model by replacing the link $0/1$ variable with a spin one $\sigma_{\alpha} = 2A_{\alpha} - 1$, carefully keeping track of all the shifts in global and local term. We refer to the works \cite{can2017critical, can2019annealed, can2022annealed} and related papers for a rigorous treatment of this class of problems. However we are interested in a slightly different setting, as the interaction matrix $\beta$ for us is indeed a generalized network adjacency matrix on the space of links, but there is no real separation between critical temperature and geometry. As a matter of fact the critical value of the $\theta$ parameter we found is a mixture of degree of the underlying regular annealed graph (the eigenvalue) with the critical temperature itself.

\subsection{Heterogeneous rank-one interaction}
\label{sec:squarelike}
As a natural generalization of the previous homogeneous model, which was given as a lamppost to highlight the main features of the method at stake, we examine a slightly more complicated, heterogeneous 1-rank model. We will show that this interaction structure already gives rise to interesting phenomena that capture the power of the single-link probability representation found in Equation \eqref{eq:probsinglelink}.  The physical content of this generalization is that genuine degree heterogeneity can arise as a spontaneously broken-symmetry phase of an otherwise homogeneous pairwise coupling, rather than being imposed by hand through node-specific local fields $\epsilon_{ij}$.
Let us consider a Hamiltonian with the general rank-one interaction matrix $\mathbf{\beta} = \lambda \ket{\omega}\bra{\omega}$

\begin{equation}
    H(\mathbf{A}) = - \sum_{i<j}\epsilon_{ij}A_{ij} - \frac{\lambda}{2V}\sum_{\substack{i<j\\k<l}} \omega_{ij}\omega_{kl} A_{ij}A_{kl}
\end{equation}

In this case, applying the general Formula \eqref{eq:probsinglelink}, the link probability would simply read
\begin{equation}
    p_{ij} = \frac{e^{\epsilon_{ij} + x^* \sqrt{\frac{\lambda}{V}} \omega_{ij}}}{1+e^{\epsilon_{ij} + x^* \sqrt{\frac{\lambda}{V}} \omega_{ij}}}
\end{equation}

What we can already see is that, according to the value that $x^*$ assumes, the model displays two different behaviors: the first one is the high temperature one, in which local link variables are the ones that explain the whole system. The other regime is the one in which the parameter $x_k^*$ is switched on and the interplay between local link parameters and the coupling structure comes into play.

\subsubsection{Heterogeneity as a low temperature phenomenon}

Let us assume then that the model is even simpler and the local variable $\epsilon_{ij} = \epsilon$ is constant  as in a standard external mean field situation. We also put ourselves in the situation of sparsity, that means the local field is large and negative. As we show in SI, this means that we can safely approximate $x_1^*=0$ as the phase parameter in the high-temperature solution. By writing the single-link probability in the form of Equation \eqref{eq:probsinglelink}, the possible values of $p_{ij}$ are of two kinds: a constant/homogeneous one in the high temperature/non effectively interacting phase, and an heterogeneous value for the other phases,   that will activate if the values of the parameters $\theta, \epsilon$ and the structure of the interaction allow to.
This can be seen also from the effective Hamiltonian of this model reading
\begin{widetext}
    \begin{equation}
    H(\mathbf{A} ,  x^*) = - \braket{\mathbf{A}|\epsilon'(x^*)} \\
    = \begin{cases}
        -   \sum_{i<j} \epsilon A_{ij} \text{ for } x^* = 0 \\
        - \sum_{i<j} \epsilon'_{ij} A_{ij} = - \sum_{i<j} \left(\epsilon + x^* \sqrt{\frac{\lambda}{V}} \omega_{ij}\right)A_{ij} \text{ when } x^*\neq 0
    \end{cases}
\end{equation}
\end{widetext}

In such a simple case already we see that the different phases describe very different models: one is a standard homogeneous Erd\"os-Renyi with probability given by $p = \frac{e^\epsilon}{1+e^\epsilon}$, and the other possible phases are those in which heterogeneity comes into place. 

Moreover, the decomposition of Equation \eqref{eq:diagonalization} can be read in two ways: if given a coupling structure one can decompose it in the spectral way, conversely one can constructively build interactions from scratch by assigning to local node variables latent parameters, that are silent in the high temperature phase but enhance the model when they activate. For example we can choose the interaction to be in such a way that $\omega_{ij} = \omega_i + \omega_j$. Then this reads as an effective Configuration Model in the low temperature phase as 

\begin{equation}
    p_{ij} = \frac{\phi_i \phi_j}{1+\phi_i \phi_j}
\end{equation}

where $\phi_i = e^{\epsilon'_i} \equiv e^{\frac{\epsilon}{2}+\sqrt{\frac{\lambda}{V}}\omega_i}$ and $\phi_j = e^{\epsilon'_j}  \equiv e^{\frac{\epsilon}{2}+ \sqrt{\frac{\lambda}{V}}\omega_j}$. The effective Hamiltonian will read indeed 

\begin{equation}
    H(\mathbf{A} ,  x^*) =  \begin{cases}
        -  \sum_{i<j} \epsilon   A_{ij} \text{ for } x^* = 0 \\
        - \sum_{i<j} \left(\epsilon'_i + \epsilon'_j \right)A_{ij} \text{ when } x^*\neq 0
    \end{cases}
\end{equation}

If we bring back the heterogeneity in the $\epsilon_{ij}$ parameter as well, assuming the same node-specific separation $\epsilon_{ij} = \epsilon_i + \epsilon_j$ we will have a family of Configuration Models, described by the probability 

\begin{equation}
    p_{ij} = \frac{  \phi_i \phi_j}{1 +  \phi_i \phi_j} 
\end{equation}

where $\phi_i = e^{\epsilon'_i} \equiv e^{\epsilon_i + \sqrt{\frac{\lambda}{V}} x^* \omega_i}$. In this case too the Hamiltonian will assume the familiar form

\begin{equation}
    H(\mathbf{A} ,  x^*) =  \begin{cases}
        - \sum_{p<q}\epsilon_{pq}\sum_{i\in p, j\in q} A_{ij} \text{ for } x^* = 0 \\
        - \sum_{i<j} \left(\epsilon'_i + \epsilon'_j \right)A_{ij} \text{ when } x^*\neq 0
    \end{cases}
\end{equation}

In all this cases the low rank coupling structure we add is the responsible for the heterogeneity of the resulting model in the low temperature phase. Perhaps the case of greatest relevance is then when a community structure is enforced in the likes of a Stochastic Block Model for the $\epsilon$ parameter, and such low rank structure is kept for the coupling. If we keep the same separation of $\omega$ the Hamiltonian will be assuming the form

\begin{widetext}
\begin{equation}
    H(\mathbf{A} ,  x^*) =  \begin{cases}
        - \sum_{p<q}\epsilon_{pq}\sum_{i\in p, j\in q} A_{ij} \text{ for } x^* = 0 \\
        - \sum_{p<q}\sum_{i\in p, j\in q} \left[\epsilon_{pq} +  x^* \sqrt{\frac{\lambda}{V}}\left( \omega_i + \omega_j \right)\right]A_{ij} \text{ when } x^*\neq 0
    \end{cases}
\end{equation}
\end{widetext}

which is a family of Stochastic Block Models that in the low temperature phases assumes the form of a Degree Corrected Stochastic Block Model.

A single rank interaction structure in the likes we analyzed in this section is clearly an ad hoc construction. In the following Sections \ref{sec:comminduced},\ref{sec:twostar} we will study more plausible couplings between links and/or nodes. Yet what we wanted to highlight with this simple addition to first order models is that they can be enhanced at will, by repeating this construction to the desired interaction rank, under the same Hamiltonian structure. Then each different phase can be seen as a stand-alone first order model, with its own parameters. Hence seemingly different Network Ensembles can be recomprised under the umbrella of a single, interacting model. It is known that Network Models can increase at will the number of parameters to better replicate given data, yet in this case we show that nestedness of Random Network Models could be a consequence of the coupling structure the networks are endowed with, and not just stand-alone first order parameters.

\subsection{Community-Induced Interactions}
\label{sec:comminduced}
We now extend the complexity of the simple models we have solved to include interacting link communities. Such communities, involving link instead of nodes, can arise for different reasons and link community detection has been already a quite active field \cite{ahn2010link, villegas2022laplacian, shang2020kcore}. We will solve for a model in which link communities are induced by the nodes' ones, but the computations can be readily generalized to the desired case where communities are defined by some other property. Among the most used (simple) models in network theory there are the Stochastic Block Model (SBM) and Core-Periphery one (CP). They both  partition the node space into different groups, called communities, $2$ for the core-periphery model and more for the block model. We want to show here that such partitions can induce quite naturally a coupling structure between links,   and if this community partitioning is reflected in the interactions, a quite complicated phase space can naturally emerge. We proceed by creating the following toy-model coupling structure:

\begin{itemize}
    \item For simplicity, let us assume there are two communities of nodes, with $N_1$ and $N_2$ nodes in each;
    \item Links between nodes of community $1$ has a certain probability $\frac{e^{\epsilon_1}}{1+ e^{\epsilon_1}}$  of existing, links between nodes of community $2$ has a probability $\frac{e^{\epsilon_2}}{1+ e^{\epsilon_2}}$ and links between the two communities has probability $\frac{e^{\epsilon_3}}{1+ e^{\epsilon_3}}$.   We can represent the structure in a more compact way through a block matrix in the form
\begin{equation}
\epsilon = \left(\begin{array}{cc}
\epsilon_1 \boldsymbol{1}_{N_1\times N_1} & \epsilon_3 \boldsymbol{1}_{N_1 \times N_2} \\
\epsilon_3 \boldsymbol{1}_{N_2 \times N_1} & \epsilon_2 \boldsymbol{1}_{N_2 \times N_2}
\end{array}\right)
\end{equation}
where $1_{N\times M}$ is the all-ones rectuangular $N \times M$ matrix;
    \item This community structure for the nodes induces a community structure for the links: the links between nodes of community $1$, that are $V_1 = N_1(N_1-1)/2$, the links between nodes of community $2$ that are $V_2 = N_2(N_2-1)/2$, and the links between nodes belonging to different communities that are $V_3 = N_1 N_2$;
    \item This  induces in turn a $3\times3$ block structure of the interaction matrix,  as if there was a Stochastic Block Model for three communities of links in the dual network in the form

\begin{equation}
\beta = \left(\begin{array}{ccc}
\beta_1 \boldsymbol{1}_{V_1\times V_1} & \beta_{12} \boldsymbol{1}_{V_1 \times V_2} & \beta_{13} \boldsymbol{1}_{V_1 \times V_3} \\
\beta_{12} \boldsymbol{1}_{V_2\times V_1} & \beta_{2} \boldsymbol{1}_{V_2 \times V_2} & \beta_{23} \boldsymbol{1}_{V_2 \times V_3} \\
\beta_{13} \boldsymbol{1}_{V_3\times V_1} & \beta_{23} \boldsymbol{1}_{V_3 \times V_2} & \beta_{3} \boldsymbol{1}_{V_3 \times V_3}
\end{array}\right)\label{eq:betainteraction}
\end{equation}
\end{itemize}

\begin{figure}[h]
    \includegraphics[width=\linewidth]{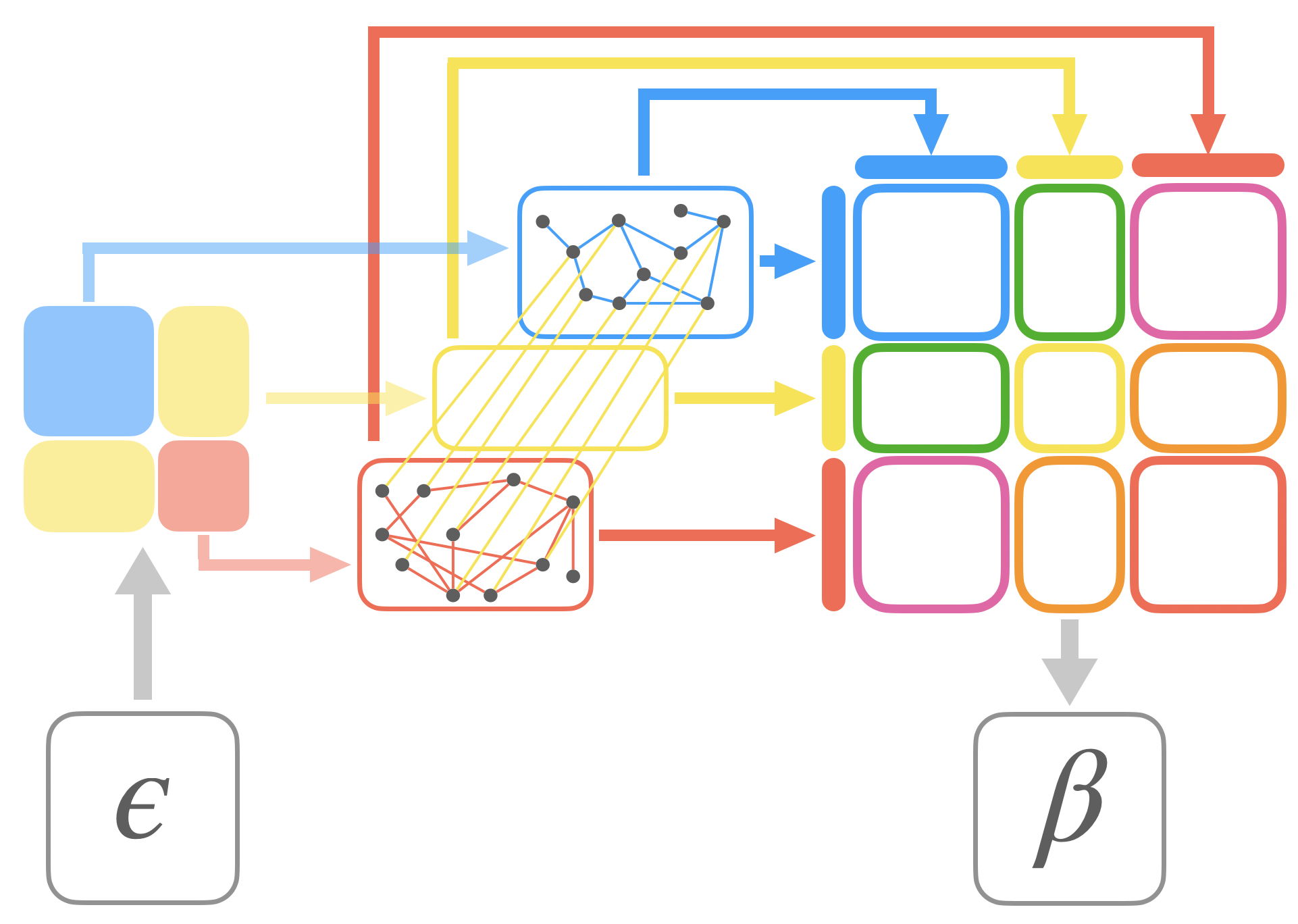}
\caption{\label{fig:communityinteraction} A pictorial representation of how the coupling matrix is induced. The two nodes community structure allows for three values of link probability: in the first community (blue), in the second community (red), across the two communities (yellow). The possible interactions between these links are then blue/blue, blue/yellow (green), blue/red (purple), and so on.}
\end{figure}

A scheme of how this community induced interaction is created is shown in Figure \ref{fig:communityinteraction}.

\begin{figure}[h!]
    \includegraphics[width=0.9\linewidth]{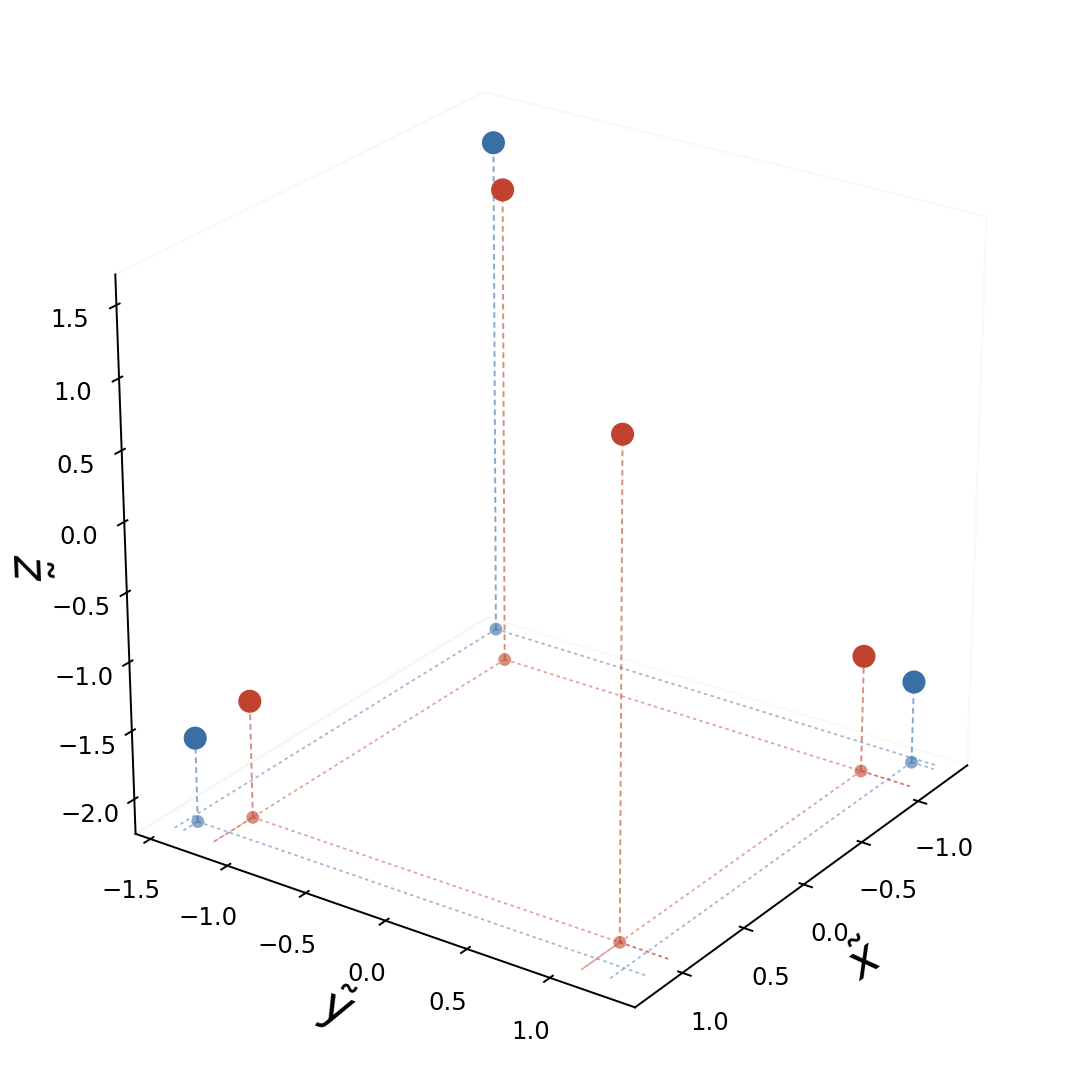}
\caption{ {Instance of a crowded phase structure in the (reduced, each variable is divided by $\sqrt{V}$) space ($\tilde x, \tilde y, \tilde z$). In blue the three critical points of the first choice of parameters, that localize in $\mathbf{x^*_1}=(1.1950,-1.3448,-1.5624)$, $\mathbf{x^*_2}=(-1.2287,1.3061,-1.5768)$, $\mathbf{x^*_3}=(-1.2677,-1.4000,1.4756)$. In red the four maximizers of the second choice of parameters, that are $\mathbf{x^*_1}=(1.0141,1.1422,1.3236)$, $\mathbf{x^*_2}=(1.0141,-1.1422,-1.3236)$, $\mathbf{x^*_3}=(-1.0141,1.1422,-1.3236)$ and $\mathbf{x^*_4} =(-1.0141,-1.1422,1.3236)$.}}\label{fig:morephases}
\end{figure}

\begin{figure}[h!]
    \includegraphics[width=0.9\linewidth]{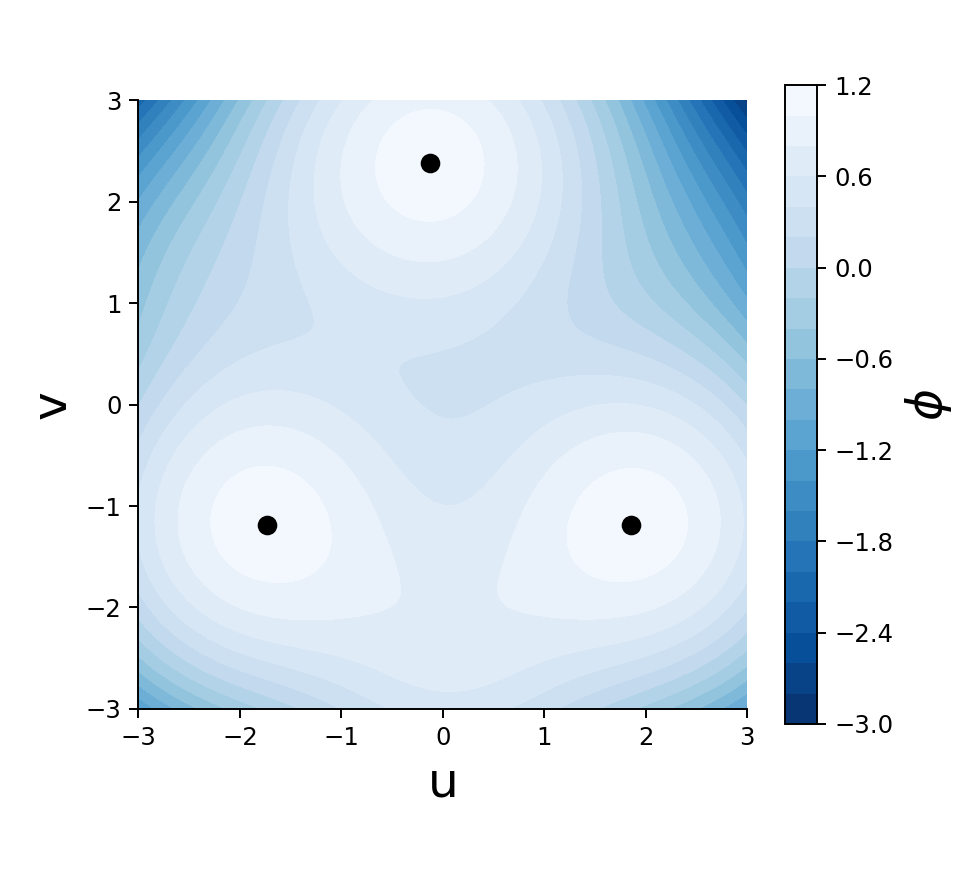}
\caption{ {Heatmap of the action $\Phi$ in the $(u,v)$ plane defined as the one passing by the three critical (dark blue) of the action of Figure \ref{fig:morephases}. The maxima attain the same value of the action, which entails true triphase coexistence. } }\label{fig:morephases2}
\end{figure}

This three by three matrix is the first example we can devise of phase space that has  {a region in the parameter space where more than two phase coexist, as we show in Figure \ref{fig:morephases}\ref{fig:morephases2}. In SI we show in detail how one can diagonalize the community induced interaction matrix \eqref{eq:betainteraction} in its three eigenmodes ($\lambda_1$, $\lambda_2$, $\lambda_3$ and respective eigenvectors $\ket{\omega_1}$, $\ket{\omega_2}$, $\ket{\omega_3}$). By computing the partition function on its saddle, we can again identify the values of $\mathbf{x^*} = (x^*, y^*, z^*)$ for which the integrand in \eqref{eq:zprod} is maximized and hence the partition function can be approximated to. In particular, for nodes in the first community, one gets the single-link probability }

 {\begin{equation}
    p_1 = \frac{e^{\epsilon_1 +  x^* \sqrt{\frac{\lambda_1}{V}}(\omega_1)_1 + y^* \sqrt{\frac{\lambda_2}{V}}(\omega_2)_1 + z^* \sqrt{\frac{\lambda_3}{V}}(\omega_3)_1}}{1+e^{\epsilon_1 +  x^* \sqrt{\frac{\lambda_1}{V}}(\omega_1)_1 + y^* \sqrt{\frac{\lambda_2}{V}}(\omega_2)_1 + z^* \sqrt{\frac{\lambda_3}{V}}(\omega_3)_1}}
\end{equation}}

 {where the specific values of $\mathbf{x^*}$ are the critical points of the action $\Phi$ for the interaction defined by $\beta$ and $\epsilon$. In Figures \ref{fig:morephases} \ref{fig:morephases2} we display that, according to different choices of the strength in the interaction matrix and the community structure one is enforcing, one can indeed identify multiple maxima in the $(x,y,z)$ space that maximizes $\Phi$, identifying regions in the parameters space where more than two phases are coexisting. In the specific case of Figures \ref{fig:morephases}, \ref{fig:morephases2}, the two communities are chosen to be equal in size $N_1=N_2$, but according to the different values one assigns to the local fields and interactions, separation into multiple critical points starts to occur. In particular, we show:}
\begin{itemize}
    \item  {In blue, the maxima of the action for the choice of local fields $\epsilon_1 = -2$, $\epsilon_2 = -1.9$, $\epsilon_3 = -1.95$, and for the interaction coefficients $\beta_1 = \beta_2 = \beta_3= 6.1$, $\beta_{12} = -1.3$, $\beta_{13}= -2.1$, $\beta_{23}= -2.7$. To further show they are actually maximal points, in Figure \ref{fig:morephases2} we also plot the heatmap in of $\Phi$ on the plane defined by them.}
    \item  {In red, the maxima of the action for a system with homogeneous local fields  $\epsilon_1 =\epsilon_2 = \epsilon_3 = -1.8$, but interaction coefficients $\beta_1 = \beta_2 = \beta_3= 15$, $\beta_{12} = -4$, $\beta_{13}= -5$, $\beta_{23}= -6.6$.}
\end{itemize}

\subsection{Node-mediated interaction}
\label{sec:twostar}
\begin{figure}[h!]
    \includegraphics[width=\linewidth]{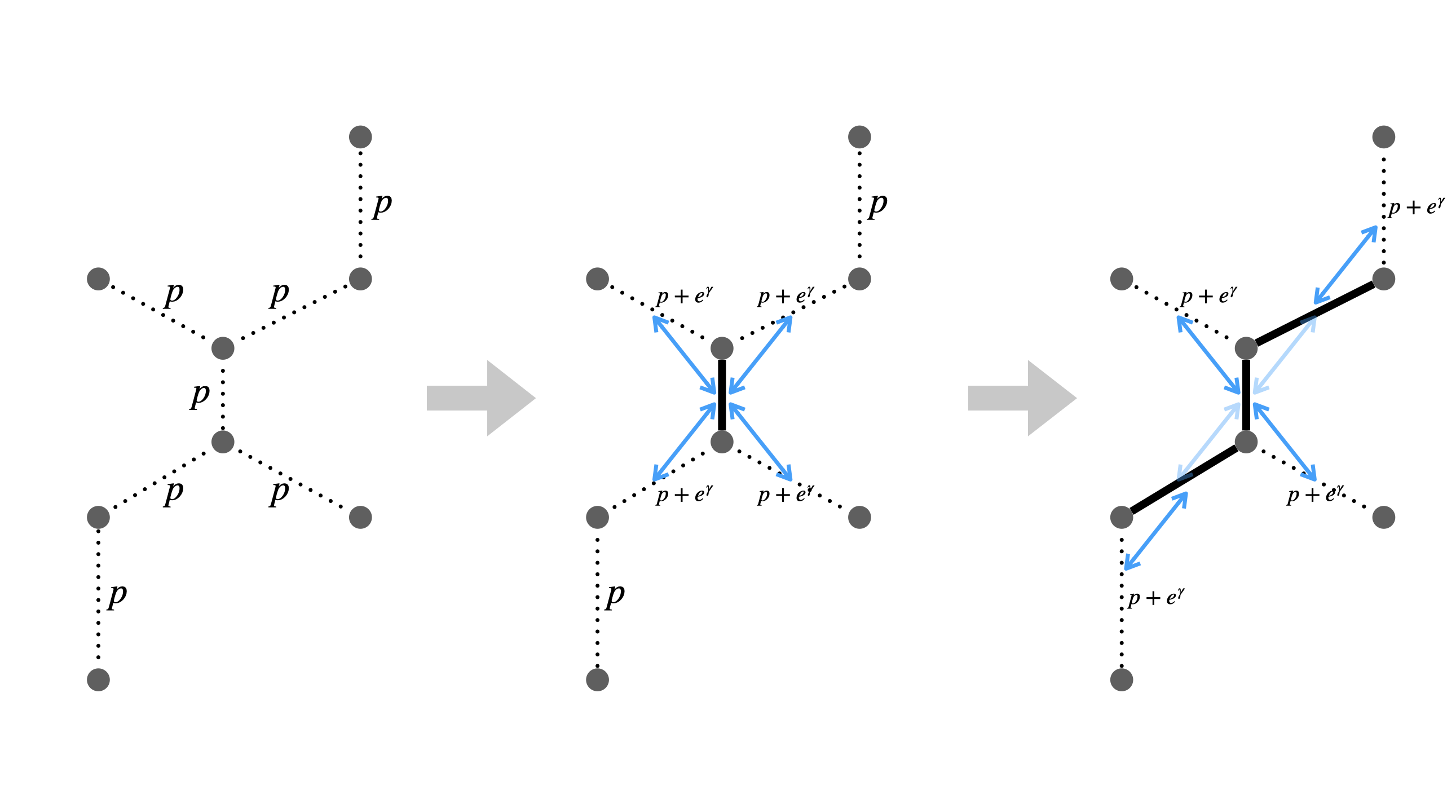}
\caption{\label{fig:twostar} A scheme of how the interaction is mediated by nodes. At first, any link has the same (low) probability $p$ to exist. If one flips up and connects two nodes, then it start interacting with the sites of the links that share one node with that link. Then more links might be promoted and propagate the interaction with their node-neighbours.}
\end{figure}

Up until now we considered models in which some kind of community structure was induced between links, and this fact was the responsible for the phase transition. Now we consider a new kind of interaction, in which links are correlated via nodes' property, namely links are correlated if they share one node in common. This structure is highly non-trivial and has been considered in the literature of ERGs as the Two Star model \cite{park2004solution, annibale2015two}. In the following we will solve it in a different fashion, making us of some results coming from algebraic coding theory. A scheme of how this interaction works is depicted in Figure \ref{fig:twostar}.

\subsubsection{The Johnson graph}
Again we assume that all $\epsilon$ are the same for simplicity, but note that once one has diagonalized the matrix of the couplings, Formula \eqref{eq:probsinglelink} allows to write the solution of a two-star model with heterogeneous degrees. Let us concentrate on the matrix then. The interaction we want to implement is such that links are correlated if they share one node. This kind of interaction can mimic for instance traffic jam propagation, where link states represent congested/uncongested streets. The probability for a street $a$ to be congested is increased if there is a congested street $b$ that shares a crossroads with $a$. We can define $\mathbf{\beta}$ as follows

\begin{equation}
    \beta_{ijkl} = \gamma \frac{V}{N} J_{(ij)(kl)}
\end{equation}
 
 where
 
\begin{equation}
    J_{(ij)(kl)} = \begin{cases}
        1 \text{ if } |(i,j)\cap(k,l)| = 1\\
        0 \text{ otherwise.}
    \end{cases}
\end{equation}
 {and $\gamma$ is the Lagrange multiplier controlling for the intensity of wedges counts.} The above matrix is actually a known object in Algebraic Coding Theory, where it is known as the adjacency matrix of the Johnson Graph $J(N, 2)$. The Johnson Graph $J(N,k)$ is a graph in which the nodes are the subsets of $\{1, 2, \ldots, N\}$ elements of dimension $k$. Edges between nodes of the Johnson Graph exist if the intersection of the two nodes has dimension $k-1$. If $k=2$ the adjacency matrix of $J(N,2)$ reduces exactly to the structure we are looking for. Moreover if one wants to treat this model as the Two-Star ERG, we can see that the Hamiltonian can be reparametrized with the choices $\frac{N\theta_1 - \theta_2}{2N} = \epsilon$ and $\theta_2 = \gamma$ and put in the more familiar form

\begin{eqnarray}
    H(A) &=& - \frac{N\theta_1 - \theta_2}{2N}\sum_{i < j} A_{ij} - \frac{\theta_2}{2 N}\sum_{i< j} \sum_{k < l}J_{ijkl} A_{ij}A_{kl} \nonumber\\
    &=& - \frac{N\theta_1 - \theta_2}{2N}\sum_{i < j} A_{ij} - \frac{\theta_2}{2 N}\sum_{i< j < k} A_{ij}A_{jk} \nonumber \\
    &=& - \theta_1 m(A) - \frac{\theta_2}{N} s(A)
\end{eqnarray}

where $m(A)= \frac{1}{2}\sum_{i<j}A_{ij} $ and $s(A) = \frac{1}{2}\sum_{i}k_i(k_i - 1)$. From the literature it is known that the Johnson Graph Adjacency matrix is full rank, and has known eigenspaces \cite{delsarte1973algebraic, vorob2020reconstruction, filmus2016orthogonal}. For our case of interest $k=2$, the eigenvalues of $\beta$ are

\begin{equation}
    \begin{cases}
        \lambda = 2N-4 \text{ with multiplicity }m_1 = 1; \\
        \mu = N-4 \text{ with multiplicity }m_2 = N-1; \\
        \nu = -2 \text{ with multiplicity }m_3 = \frac{(N-2)(N-3)}{2};
    \end{cases}
\end{equation}

 {We can then write the partition function in the form of Eq. \eqref{eq:zprod}, where the action $\Phi$ is}
 {
\begin{widetext}
    \begin{equation}
        \Phi(\mathbf{x}) = -\sum_{k=1}^V \frac{x_k^2}{2} + \sum_{a=1}^V \ln\left( 1 + e^{\epsilon + x_1 \sqrt{\frac{\gamma \lambda}{N}} (\omega_1)_a + \sum_{k=2}^N x_k \sqrt{\frac{\gamma \mu}{N}} (\omega_k)_a + \sum_{k=N+1}^V x_k \sqrt{\frac{\gamma \nu}{N}} (\omega_k)_a} \right)
    \end{equation}
\end{widetext}}

where all the $V$ modes are arranged in the three different contributions given by the different eigenspaces.

\subsubsection{The large $N$ limit}
We now want compute the terms of the partition function in the large $N,V$ limit.  {Again, a saddle point evaluation argument allows us to compute the partition function by approximating its value with the value of the integrand in its maximal points. The conditions for which this happens are the critical points of $\Phi$, namely those for which}

 {
\begin{equation}
    x_1^* = \sqrt{\frac{\gamma\lambda}{N}}p(\mathbf{x^*}) \sum_{a=1}^V (\omega_1)_a
\end{equation}}
 {
\begin{equation}
    x_k^* = \sqrt{\frac{\gamma\mu}{N}}p(\mathbf{x^*}) \sum_{a=1}^V (\omega_k)_a \quad k = 2, 3, \ldots N
\end{equation}}
 {
\begin{equation}
    x_k^* = \sqrt{\frac{\gamma\nu}{N}}p(\mathbf{x^*}) \sum_{a=1}^V (\omega_k)_a \quad k = N+1, \ldots V.
\end{equation}}

 {A standard result in algebraic graph theory comes in our help in this case: as the Johnson graph its regular, we immediately know that its main eigenvector has constant, positive entries, and as each other eigenvector is orthogonal to it, all other eigenvectors' entries identically sum to zero. This immediately fixes the values of all $x_k^*$ to zero with the exception of $x_1^*$, that might still assume different values. With this in mind, we can evaluate the action into its critical point, yielding }

 {\begin{equation}
    \Phi(\mathbf{x^*}) = - \frac{(x_1^*)^2}{2}+ V \ln \left(1 + e^{\epsilon + x_1^* \sqrt{\frac{\gamma \lambda}{N}}\frac{1}{\sqrt{V}}}\right)
\end{equation}}

 {and by substituting $x_1^* = \sqrt{V}y^*$ we can easily compute the single link partition function}

 {\begin{equation}
    z = Z^{\frac{1}{V}}  \simeq e^{-\frac{(y^*)^2}{2}}\left(1 + e^{\epsilon + y^*\sqrt{\frac{\gamma \lambda}{N}}}\right)
\end{equation}}

 {that has the very same structure (although for different parameters) of eq. \eqref{eq:z_homo}. The single link probability as well can be shown to have the same structure of the Homogenous case, namely}

 {\begin{equation}
    p_{eff} = \frac{e^{\epsilon + y^* \sqrt{\gamma\frac{2N-4}{N}}}}{1 + e^{\epsilon + y^* \sqrt{\gamma\frac{2N-4}{N}}}} \simeq \frac{e^{\epsilon + y^* \sqrt{2\gamma}}}{1 + e^{\epsilon + y^* \sqrt{2 \gamma}}}
\end{equation}}

\begin{figure}[h]
    \includegraphics[width=\linewidth]{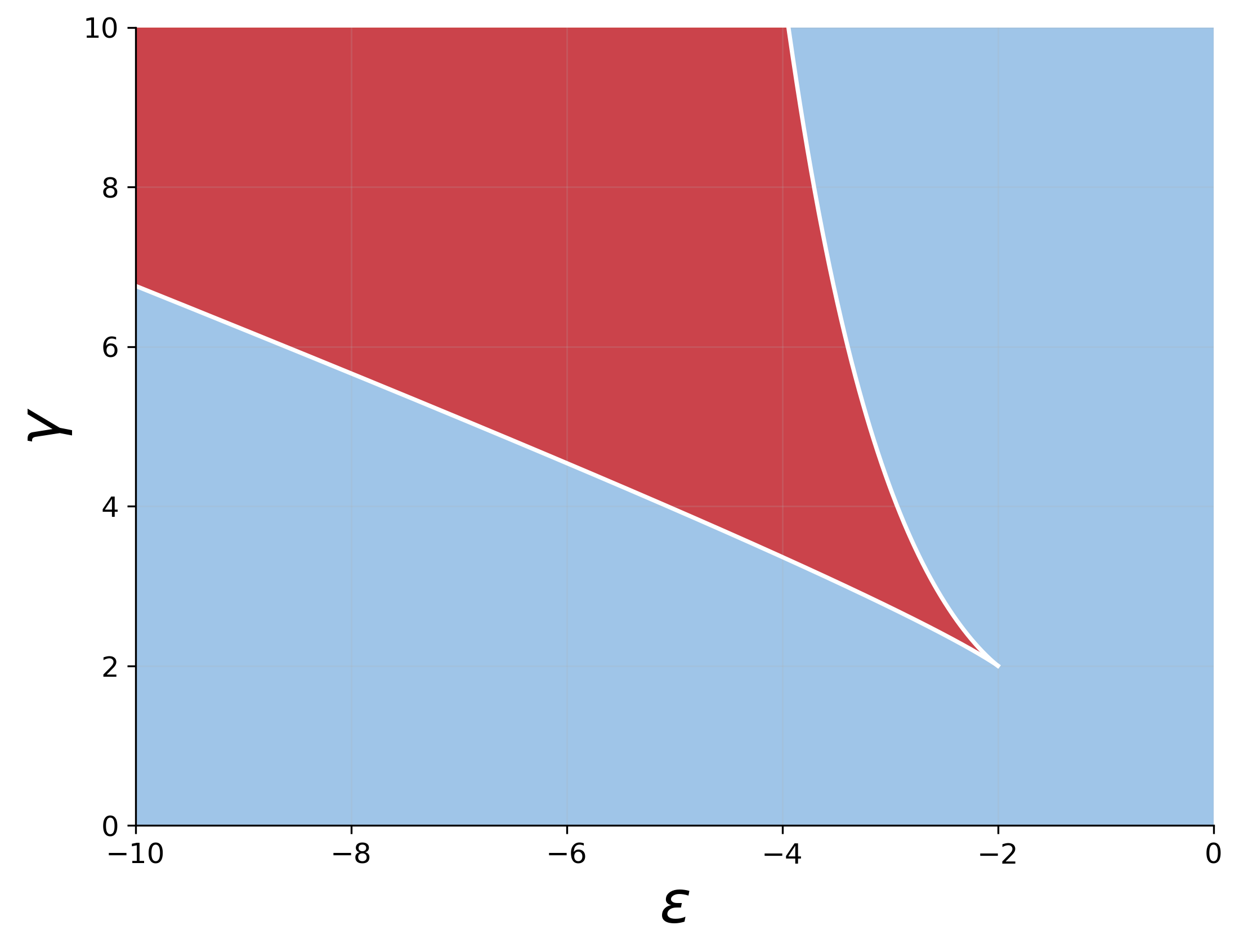}
\caption{\label{fig:PD_twostar} Phase Diagram of the Node-mediated Interaction model. In red the region where multiple phases are present,  in blue the region where only one phase is possible.}
\end{figure}

\begin{figure}[h]
    \includegraphics[width=\linewidth]{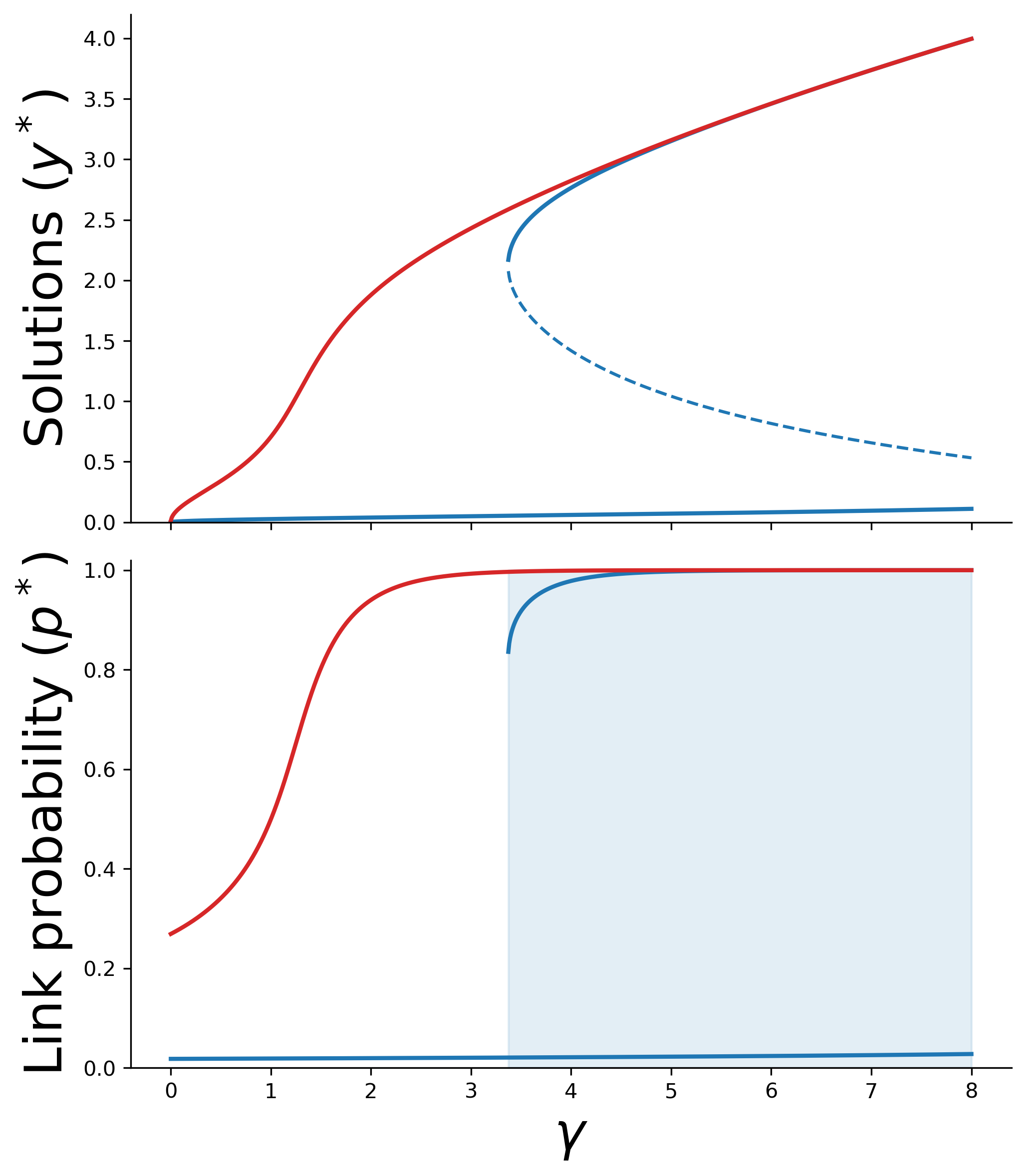}
\caption{\label{fig:plink_twostar} Single-link  {partition function and} probability in the range $\gamma =[0,8]$,  shaded region indicate where two solutions are possible,  {dashed line indicate critical, minimum point}.}
\end{figure}

In Figures \ref{fig:PD_twostar} and \ref{fig:plink_twostar} we show the Phase Diagram and the link probability of the model for a certain range of parameters, which both resemble the simple, homogeneous case.

 One of the power of the diagonalization method we use is that once one has an orthogonal basis for the interaction matrix, any dual graph on links that has the same algebra (\textit{i.e.} the same eigenvectors) can be used as a coupling structure and straightforwardly solved. Indeed, the only thing that one need to replace in Equations \eqref{eq:zsinglelink} \eqref{eq:probsinglelink} is the values of the of the new eigenvalues, as it is the only thing that would change. For example, in the case of the Kneser graphs \cite{kneser1955aufgabe360, lovasz1978kneser} (that are the complement of the Johnson graphs analyzed above), we can easily determine the fact that the phase diagram would not be much different from the one of Figure \ref{fig:PD_twostar}, even though the actual values in which the partition function splits would be different, as the eigenvalues of the two graphs are.

\section{On the Mean Field Approximation and Phase Structure}
\label{sec:meanfield}

Before the final remarks, we want to comment on the unexpected similarity between the node-mediated interaction model \ref{sec:twostar} and the simple, homogeneous one \ref{sec:homo}. One would expect that with such a non-trivial correlation structure, the mean field computation, which is essentially the model of Section \ref{sec:homo} with the right scalings for the parameters, would be a wrong approximation of the system. However that is not the case, and they basically coincide in the large $N$ limit.  Explicitly, the correspondence follows from  {the fact that the regularity of the interaction kills the chance the critical points of the action have to be other than zero. This is expected to happen also for non exactly regular interactions, that nonetheless display a correlation that is homogeneous spread out over all links. This is clear if one looks at the stationarity equations \eqref{eq:stationar} defining the critical points of the action. }

 {It is worth making explicit why this happens, since the argument
generalizes well beyond the two models compared here. From
Eq.~\eqref{eq:stationar}, a nonzero solution $x_k^*\neq 0$ for a mode $k$ requires the
scalar product $\langle\omega_k|p(\mathbf{x}^*)\rangle$ to be nonzero,
weighted by $\sqrt{\lambda_k/V}$. The regular case is special precisely
because algebraic graph theory fixes the \emph{shape} of the
non-principal eigenvectors independently of the interaction strength:
for a regular dual graph, orthogonality to the (constant,
Perron--Frobenius) principal eigenvector forces every other eigenvector
to sum identically to zero over its $V$ entries, so that any
\emph{homogeneous} contribution to $p(\mathbf{x}^*)$ is automatically
annihilated in the scalar product, however large $\lambda_k$ is. What
survives is only the principal mode, whose eigenvector has constant
sign and therefore cannot be orthogonal to an all-positive probability vector.
The net effect is that a regular, or near-regular, interaction
structure behaves as an effective one-dimensional (mean-field) problem
no matter how large its rank, which is the precise sense in which the
mean-field approximation is ``good'' here: it is not that the
corrections are small in some perturbative sense, but that the
corrections are identically absent at the level of the saddle-point
equations themselves.}

 {If one wishes to observe a phase separation
structure arising from non-principal eigenmodes, then a simple rule of
thumb can be written as a two-criteria checklist:}

 {\begin{enumerate}
\item[(i)] \emph{Extensive eigenvalue}: $\lambda_k/V$ tends to a
finite, nonzero constant (or $\lambda_k/N$ if the growth of the largest
eigenvalue is of order $N$, as in the Johnson graph case). This
condition is best understood as the network-theoretic analogue of the
Baik--Ben Arous--P\'ech\'e (BBP) transition for spiked random
matrices~\cite{BaikBenArousPeche2005}: in that setting, a finite-rank
perturbation of strength $\alpha$ added to a large random (or, here,
structured) matrix produces a genuine \emph{outlier} eigenvalue (one
that detaches from the bulk of the spectrum and whose eigenvector
acquires a nontrivial, delocalized overlap with the signal) only once
$\alpha$ crosses a critical threshold set by the edge of the bulk
spectral density; below threshold, the perturbation is absorbed into
the bulk and leaves no macroscopic trace. Our extensivity condition
plays exactly this role for the dual link-coupling matrix $\beta$: a
mode $\lambda_k$ that scales sub-extensively with $V$ (or $N$, as
appropriate) is washed out in the $V\to\infty$ limit used to define the
saddle point.
\item[(ii)] \emph{Non-orthogonality between probabilities and
eigenvectors}, i.e.\ $\langle\omega_k|p(\mathbf{x}^*)\rangle\neq 0$ for
all $k>1$. As noted above, this is a sum of $V$ terms with both
positive and negative contributions and a normalized eigenvector, so
each term is individually suppressed by a factor $1/\sqrt{V}$; the
condition therefore asks that these $O(1/\sqrt{V})$ contributions do
not cancel systematically, which is precisely what happens whenever
$\omega_k$ carries a \emph{extensiviely localized} or \emph{community-like}
structure \textit{i.e.} an eigenvector whose entries are not spread with
zero mean across all links, but instead concentrate with a consistent
sign on an extensive subset of links. This connects naturally to
results on the detectability of community structure via spectral
methods, where a community-induced eigenvalue must separate from the
bulk of the spectrum by an amount related to the community strength
before the corresponding (delocalized) eigenvector becomes
informative~\cite{NadakuditiNewman2012}, exactly mirroring the
criterion above, and is also structurally the same condition used to
assess whether a spike is detectable in the spiked Wigner/covariance
models of random matrix
theory~\cite{BaikBenArousPeche2005,Noiry2019}.
\end{enumerate}}

 {Together, the two criteria say something quite intuitive: only
interactions that are simultaneously (i) \emph{strong enough, relative
to system size}, and (ii) \emph{structured enough that this strength
does not average itself away over the link population} can drive the
model away from its mean-field, first-mode-only behaviour.}

\begin{figure}[h!]
    \includegraphics[width=0.8\linewidth]{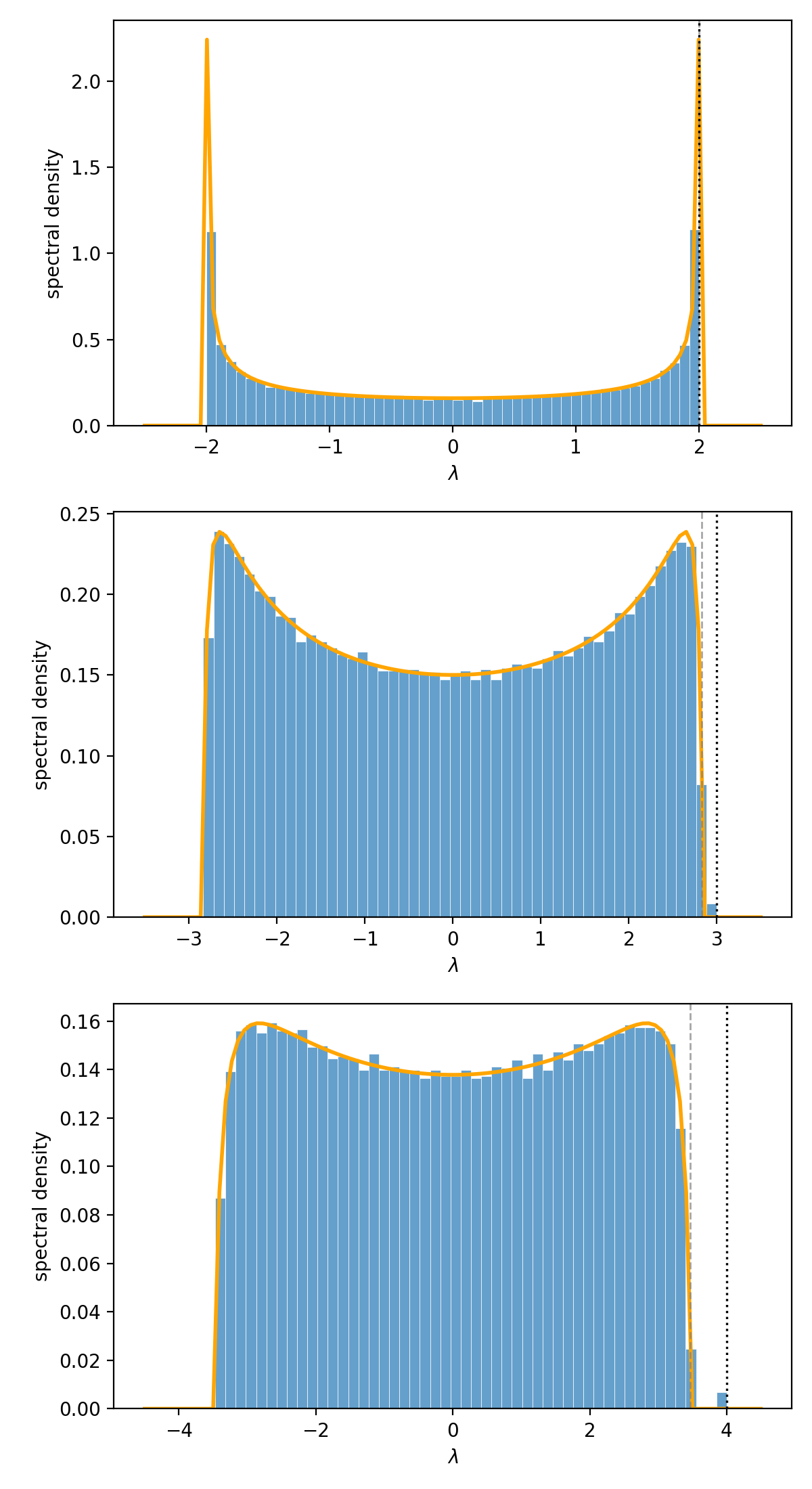}
\caption{\label{fig:regularspectra} Spectrum of the adjacency matrix of a random regular graph with $1000$ nodes. In orange the Kensten-McKay law describing the distribution of the bulk of the spectrum. From top to bottom the degree $k$ is $2$, $3$, $4$, and we can see that the outlier starts to appear from the second figure only.}
\end{figure}

 {This heuristic for the outliers to be responsible for the phase
transitions can be further argued for in the case of the regular random
graph, as we hinted at in Section~\ref{sec:homo}. Indeed, looking at the spectra
of regular random graphs, the first one in which a separation of the
spectrum appears is when the degree is 3, as can be seen in Figure~\ref{fig:regularspectra},
consistent with known results on the detectability threshold for
spiked regular graph ensembles~\cite{NadakuditiNewman2012} and
with the general BBP phenomenology.}

 {Finally, we note a practical consequence of framing the criteria this
way: since the BBP-type threshold is a property of the \emph{bulk
spectral density} of the coupling matrix $\beta$ rather than of any
single eigenvalue in isolation, the same two-criteria checklist can, in
principle, be used operationally on empirical data. Given a fitted or
estimated interaction matrix $\beta$, one can compare its empirical
spectral density against standard random-matrix-theory null
distributions for the bulk (e.g.\ the Kesten--McKay law used in
Figure~\ref{fig:regularspectra} for regular graphs, or its generalizations for irregular
degree sequences) to flag outlier eigenvalues as candidate drivers of
phase separation, and then check condition~(ii) directly on the
corresponding eigenvectors.}

\section{Final Remarks}

The solution of Random Network Models is a fundamental, analytical tool to tackle many among the problems in the science of Complex Systems. In this paper we have shown how to solve the models that mimic a link-link coupling of real world systems. The power of our approach lies in the fact that is quite universal. Using the general Formula \eqref{eq:probsinglelink}, heterogeneity in local features can be added at will while still retaining the same coupling structure. On the condition of having a viable way to decompose the interaction structure, we have also shown that Equation \eqref{eq:zprod} can provide an explicit solution of the partition function and the phase structure as well.
 An immediate practical consequence of this explicit solution is that it makes maximum
likelihood estimation directly accessible for this class of second order models. Since the partition function of Equation \eqref{eq:zprod} and the single-link probability of Equation \eqref{eq:probsinglelink} are available in closed, or at least saddle-point-exact, form, the likelihood of an observed network under any of the models we solve can be written down and maximized without resorting to the Markov Chain Monte Carlo or pseudo-likelihood approximations ordinarily required for pairwise-dependent exponential random graph models \cite{Snijders2002, HunterHandcock2006}, and without incurring the model-degeneracy pathologies such approximate schemes are known to suffer from \cite{Handcock2003}. We regard this as one of the more promising practical directions opened up by the present work: using the class of second order models solved here as tractable, exactly-normalizable null models for network reconstruction and
hypothesis testing, in the same spirit already established for first order maximum-entropy models\cite{squartini2017maximum}.

 {Moreover, in the various cases that we explored, we have seen that the
phase structure can become easily crowded as the phase landscape can be
quite hilly. Higher order features of networks are increasingly thought
to be key features of complex
systems~\cite{Battiston2020,Boccaletti2023}, and the present work
constitutes a little brick in the wall of understanding and solving
coupled, complex structures~\cite{shang2022threebody}. But we have also shown in
Section~\ref{sec:squarelike} how seemingly higher order structures can be effectively
reduced to lower order network models and be, in principle,
indistinguishable from them, a reduction that is only possible because
the interaction we study remains, formally, a pairwise coupling between
edges rather than an irreducible higher-body term.
}

 {Lastly, the study of even higher order structure has still many more
steps to undertake, and we believe some of the methods developed here,
and properly generalized, could be of help in that direction of
research, for example as null models to distinguish between true higher
order interaction and the one just described by link-link coupling that
is extensively dealt with in the present work. A particularly concrete
direction suggested by Section~\ref{sec:meanfield} is to use the extensive-eigenvalue /
non-orthogonality criterion as a diagnostic: given an empirical network,
one could fit the best second order (pairwise edge-coupled) null model,
and treat the spectral outliers of the fitted interaction matrix
$\beta$ that fail our criteria as evidence that a genuinely higher
order description is required, rather than an artifact of under-fitting
a first order model.}

\begin{acknowledgments}
This work is supported by the Dutch Econophysics Foundation (Stichting Econophysics Leiden, the Netherlands) and by the European Union - NextGenerationEU - National Recovery and Resilience Plan (Piano Nazionale di Ripresa e Resilienza, PNRR), projects ``SoBigData.it - Strengthening the Italian RI for Social Mining and Big Data Analytics'', Grant IR0000013 (n. 3264, 28/12/2021) (\url{https://pnrr.sobigdata.it/}), and ``Reconstruction, Resilience and Recovery of Socio-Economic Networks'' RECON-NET EP\_FAIR\_005 - PE0000013 ``FAIR'' - PNRR M4C2 Investment 1.3. Work by Alessio, Catanzaro is supported by \#NEXTGENERATIONEU (NGEU) and funded by the Italian Ministry of University and Research (MUR), National Recovery and Resilience Plan (NRRP), project MNESYS (PE0000006) – A Multiscale integrated approach to the study of the nervous system in health and disease (DN. 1553 11.10.2022).
\end{acknowledgments}

\bibliography{apssamp}

%
%
\clearpage
\newpage
\setcounter{equation}{0}

\setcounter{figure}{0}
\setcounter{table}{0}
\setcounter{page}{1}
\setcounter{section}{0}

\renewcommand{\theequation}{S\arabic{equation}}
\renewcommand{\thetable}{S\arabic{table}}
\renewcommand{\thefigure}{S\arabic{figure}}
\renewcommand{\thesection}{S.\Roman{section}} 
\renewcommand{\thesubsection}{\thesection.\roman{subsection}}

\onecolumngrid
{\center
\textbf{SUPPLEMENTARY INFORMATION}\\
$\quad$\\
accompanying the paper\\
\emph{``A General Solution for Network Models with Pairwise Edge Coupling''}\\
by A. Catanzaro, Subodh P. Patil and D. Garlaschelli\\
$\quad$\\
$\quad$\\
}

\section{Representation of a function of link variables}
\label{app:anyrandom}
The results in this first Section are consequences of the Relation \eqref{eq:commutA}. The first thing to realize is that any function of variables with such a relation is actually a multilinear polynomial of the variables $\{A_{ij}\}$. The intuition for this comes from a similar computation performed for Grassmannian variables (the relation between link $(0/1)$ and spin $(-1,1)$ variables is more profound than just an analogy, as a matter of fact the mapping $\sigma_{ij} = 2A_{ij} -1$ exists and allow to frame one model into the language of the other and viceversa). 

We suppose the probability function, and all the functions we take into account are nice enough for the following computations. Then we can expand a function of the relabeled link variables $A_1, A_2, \ldots , A_V$ in Maclaurin series

\begin{multline}
    f(A_1, \ldots ) = \sum_{\alpha =0}^\infty \sum_{\beta =0}^\infty \ldots \sum_{\omega =0}^\infty \frac{A_1^\alpha A_2^\beta \ldots A_V^\omega}{\alpha! \beta! \ldots \omega!} \left. \frac{\partial^{\alpha + \beta + \ldots + \omega }}{\partial A_1^\alpha \partial A_2^\beta \ldots \partial A_V^\omega}f(A_1,\ldots)\right|_{0}  \\
    = \sum_{\beta =0}^\infty \ldots \sum_{\omega =0}^\infty \frac{A_2^\beta \ldots A_V^\omega}{\beta! \ldots \omega!} \sum_{\alpha = 0}^\infty \frac{A_1^\alpha}{\alpha!} \left. \frac{\partial^{\alpha + \beta + \ldots + \omega }}{\partial A_1^\alpha \partial A_2^\beta \ldots \partial A_V^\omega}f(A_1,\ldots)\right|_{0}  \\
    = \sum_{\beta =0}^\infty \ldots \sum_{\omega =0}^\infty \frac{A_2^\beta \ldots A_V^\omega}{\beta! \ldots \omega!} \left[ \left. \frac{\partial^{\beta + \ldots + \omega }}{\partial A_2^\beta \ldots \partial A_V^\omega}f(A_1, \ldots )\right|_{0} + \sum_{\alpha = 1}^\infty \frac{A_1^\alpha}{\alpha!} \left. \frac{\partial^{\alpha + \beta + \ldots + \omega }}{\partial A_1^\alpha \partial A_2^\beta \ldots \partial A_V^\omega}f(A_1, \ldots )\right|_{0}\right] \\
    = \sum_{\beta =0}^\infty \ldots \sum_{\omega =0}^\infty \frac{A_2^\beta \ldots A_V^\omega}{\beta! \ldots \omega!} \left[ \underbrace{ \left.\frac{\partial^{\beta + \ldots + \omega }}{\partial A_2^\beta \ldots \partial A_V^\omega}f(A_1, \ldots )\right|_{0}}_{C_0} + A_1 \underbrace{ \sum_{\alpha = 1}^\infty \frac{1}{\alpha!} \left. \frac{\partial^{\alpha + \beta + \ldots + \omega }}{\partial A_1^\alpha \partial A_2^\beta \ldots \partial A_V^\omega}f(A_1, \ldots)\right|_{0} }_{C_1}\right]
\end{multline}

where we explicited the linear dependence from $A_1$. If we do the same for $A_2$ we get

\begin{multline}
    \sum_{\gamma =0}^\infty \ldots \sum_{\omega =0}^\infty \frac{A_3^\gamma \ldots A_V^\omega}{\gamma! \ldots \omega!} \sum_{\beta=0}^\infty \frac{A_2^\beta}{\beta!} \left[ \left. \frac{\partial^{\beta + \ldots + \omega }}{\partial A_2^\beta \ldots \partial A_V^\omega}f(A_1, \ldots )\right|_{0} + A_1 \sum_{\alpha = 1}^\infty \frac{1}{\alpha!} \left. \frac{\partial^{\alpha + \beta + \ldots + \omega }}{\partial A_1^\alpha \partial A_2^\beta \ldots \partial A_V^\omega}f(A_1, \ldots)\right|_{0}\right]  \\
    = \sum_{\gamma =0}^\infty \ldots \sum_{\omega =0}^\infty \frac{A_3^\gamma \ldots A_V^\omega}{\gamma! \ldots \omega!} \left[ \left[ \left. \frac{\partial^{\gamma + \ldots + \omega }}{\partial A_3^\gamma \ldots \partial A_V^\omega}f(A_1, \ldots )\right|_{0} + A_1 \sum_{\alpha = 1}^\infty \frac{1}{\alpha!} \left. \frac{\partial^{\alpha + \gamma + \ldots + \omega }}{\partial A_1^\alpha \partial A_3^\gamma \ldots \partial A_V^\omega}f(A_1, \ldots)\right|_{0}\right] \right. \\
    \left. +\sum_{\beta = 1}^\infty \frac{A_2^\beta}{\beta!}\left[ \left. \frac{\partial^{\beta + \ldots + \omega }}{\partial A_2^\beta \ldots \partial A_V^\omega}f(A_1, \ldots )\right|_{0} + A_1 \sum_{\alpha = 1}^\infty \frac{1}{\alpha!} \left. \frac{\partial^{\alpha + \beta + \ldots + \omega }}{\partial A_1^\alpha \partial A_2^\beta \ldots \partial A_V^\omega}f(A_1, \ldots)\right|_{0}\right] \right]  \\
    = \sum_{\gamma =0}^\infty \ldots \sum_{\omega =0}^\infty \frac{A_3^\gamma \ldots A_V^\omega}{\gamma! \ldots \omega!} \left[ \underbrace{\left. \frac{\partial^{\gamma + \ldots + \omega }}{\partial A_3^\gamma \ldots \partial A_V^\omega}f(A_1, \ldots )\right|_{0}}_{C_0} + A_1 \underbrace{\sum_{\alpha = 1}^\infty \frac{1}{\alpha!} \left. \frac{\partial^{\alpha + \gamma + \ldots + \omega }}{\partial A_1^\alpha \partial A_3^\gamma \ldots \partial A_V^\omega}f(A_1, \ldots)\right|_{0}}_{C_1}  \right. \\
    + \left. A_2 \underbrace{ \sum_{\beta=1}^\infty \frac{1}{\beta!} \left. \frac{\partial^{\beta + \ldots + \omega }}{\partial A_2^\beta \ldots \partial A_V^\omega}f(A_1, \ldots )\right|_{0}}_{C_2} + A_1 A_2 \underbrace{\sum_{\alpha=1}^\infty \sum_{\beta=1}^\infty \frac{1}{\alpha! \beta!} \left. \frac{\partial^{\alpha + \beta + \ldots + \omega }}{\partial A_1^\alpha \partial A_2^\beta \ldots \partial A_V^\omega}f(A_1, \ldots)\right|_{0} }_{C_{12}}\right]
\end{multline}

so we can see that by iterating this procedure that splits the sum to highlight the reducible powers of the variables, we can prove that any function of these variables is indeed a multilinear polynomial of the $A_1, A_2, \ldots$. As a consequence, any function of these variables can be expressed in the form

\begin{multline}
    f(A) = f^{(0)} + \sum_{i<j} f_{ij}^{(1)} A_{ij} + \sum_{i<j}\sum_{k<l} f_{ijkl}^{(2)} A_{ij}A_{kl} + \sum_{i<j}\sum_{k<l}\sum_{m<n}f_{ijklmn}^{(3)} A_{ij}A_{kl}A_{mn}   \\
    + \sum_{i<j}\sum_{k<l}\sum_{m<n}\sum_{o<p}f_{ijklmnop}^{(4)} A_{ij}A_{kl}A_{mn}A_{op} + \ldots \label{eq:probexpanded}.
\end{multline}

In particular this is true for the Hamiltonian function, which then assumes the form of Equation \eqref{eq:multilinearexpression}. If one wants to express a generic $P(A)$ in the ERG form \eqref{eq:ERG}, then one has to expand $P(A)$ as we just shown. Then also the function $\frac{e^{-H(A)}}{Z}$ can be easily expanded as

\begin{multline}
    \frac{e^{-H(A)}}{Z} = \frac{1}{Z}  \left(\prod_{i<j} e^{h_{ij}A_{ij}} \right) \left(\prod_{\substack{i < j \\ k<l}} e^{h_{ijkl}A_{ij}A_{kl}}\right)\left( \prod_{\substack{i < j \\ k<l \\ m< n}} e^{h_{ijklmn}A_{ij}A_{kl}A_{mn}}\right) \ldots  \\
    = \frac{1}{Z}\left(\prod_{i<j} \sum_{\alpha = 0}^{\infty} \frac{A_{ij}^\alpha h_{ij}^\alpha}{\alpha!} \right)\left(\prod_{\substack{i < j \\ k<l}} \sum_{\alpha = 0}^{\infty} \frac{A_{ij}^\alpha A_{kl}^\alpha h_{ijkl}^\alpha}{\alpha!} \right) \left(\prod_{\substack{i < j \\ k<l \\ m<n}} \sum_{\alpha = 0}^{\infty} \frac{A_{ij}^\alpha A_{kl}^\alpha A_{mn}^\alpha h_{ijklmn}^\alpha}{\alpha!} \right) \ldots  \\
    = \frac{1}{Z} \left(\prod_{i<j}\left[1 + A_{ij}\sum_{\alpha = 1}^{\infty} \frac{h_{ij}^\alpha}{\alpha!}\right] \right) \left(\prod_{\substack{i < j \\ k<l}}\left[1 + A_{ij}A_{kl}\sum_{\alpha = 1}^{\infty} \frac{h_{ijkl}^\alpha}{\alpha!}\right] \right) \left(\prod_{\substack{i < j \\ k<l \\ m<n}}\left[1 + A_{ij}A_{kl}A_{mn}\sum_{\alpha = 1}^{\infty} \frac{h_{ijklmn}^\alpha}{\alpha!}\right] \right)\ldots  \\
    = \frac{1}{Z} \left(\prod_{i<j}\left[1 + A_{ij} \underbrace{\left(e^{h_{ij}-1}\right)}_{K_{ij}}\right] \right) \left(\prod_{\substack{i < j \\ k<l}}\left[1 + A_{ij}A_{kl} \underbrace{\left(e^{h_{ijkl}-1}\right)}_{K_{ijkl}}\right] \right) \left(\prod_{\substack{i < j \\ k<l \\ m<n}}\left[1 + A_{ij}A_{kl}A_{mn} \underbrace{\left(e^{h_{ijklmn}-1}\right)}_{K_{ijklmn}}\right] \right)\ldots
\end{multline}

Again, we can identify this function as a multivariate, multilinear polynomial in the variables $\{A_{ij}\}$. One can then identify term by term the coefficients of the two polynomials and solve the systems for the $h_{ij}, h_{ijkl}, \ldots$ to express any RNM in the form \eqref{eq:ERG}.

\section{BraKet Hamiltonian}
\label{app:setupH}

We have to maipulate the second order term of Equation \eqref{eq:multilinearexpression}. The term $h^0$ can obviously be ignored as it is just a multiplicative constant in front of the partition function. We are left with

\begin{equation}
    H(A) = - \theta_1 \sum_{i< j} h^{(1)}_{ij} A_{ij} - \theta_2 \sum_{\substack{i < j, k < l}} h^{(2)}_{ijkl} A_{ij} A_{kl}\label{eq:Hgen}
\end{equation}

First and foremost we change the labels of the indices to get a more manageable expression, since we see that the sums can be expressed in the couples of indices $(i,j)$ which we order with index $a$ going from $1$ to $V=\frac{N(N-1)}{2}$ (without self-loops). We then express the interaction coefficient in order to scale with $V$ as $\theta_2 h^{(2)}_{ijkl} = \frac{\beta_{ijkl}}{2V}$.

The new Hamiltonian becomes

\begin{equation}
    H(A) = -\theta_1 \sum_{a=1}^{V}h^{(1)}_{a}A_{a} - \frac{1}{2V}\sum_{\substack{a\neq b}} ^{V}\beta_{ab} A_{a} A_{b}
\end{equation}

we further manipulate this object to have a compact term to operate with. First we see that 

\begin{equation}
    \frac{1}{2}\sum_{\substack{a\neq b}} ^{V}\beta_{ab} A_{a} A_{b} = \frac{1}{2V}\sum_{a,b} ^{V}\beta_{ab} A_{a} A_{b} - \frac{1}{2V} \sum_{a} \beta_{aa}A_{a}  = \frac{1}{2V}\braket{A|\beta|A} - \frac{1}{2V}\braket{A|\Vec{\beta}}
\end{equation}

where $\Vec{\beta}^T = (\beta_{11}, \beta_{22}, \cdots, \beta_{VV})$. 

Then we write also the first term in the braket notation so that the whole Hamiltonian is in the form

\begin{equation}
    H(A) = -  \sum_{a}^{V}A_{a}  \left(\theta_1 h^{(1)}_{a}+ \frac{\beta_{aa}}{2V}\right) - \frac{1}{2V}\sum_{a,b} ^{V}\beta_{ab} A_{a} A_{b} = -\braket{A|\epsilon} - \frac{1}{2V}\braket{A|\beta|A}
\end{equation}

where $\beta$ is a matrix  that couple the links between them selves.
This matrix is an adjacency matrix (either expected or realized) in disguise, over the network of links of the actual graph (dual or line graph in the literature), and we will make use of the properties of such a matrix to derive the behaviour of the models.

We assume the $\beta$ matrix is symmetric (as the coupling between links is). Then it admits a spectral representation 

\begin{equation}
    \beta = \sum_{k=1}^r \lambda_{k}\ket{\omega_k}\bra{\omega_k}
\end{equation}

where $r$ is the rank of the matrix, $\lambda_k$ are the eigenvalues and $\ket{\omega_k}$ their respective orthonormal eigenvectors. We also split the positive spectrum part from the negative one. Let then $r_+$ be the number of positive eigenvalues, (including their multiplicities) and $r-r_+$ the number of the negative ones.

The Hamiltonian can be hence written as

\begin{equation}
    H(A) = -\braket{A|\epsilon} - \sum_{k=1}^{r_+} \frac{\lambda_k}{2V} \braket{A|\omega_k}^2 - \sum_{k=r_+ +1}^r \frac{\lambda_k}{2V} \braket{A|\omega_k}^2 \label{eq:Hsepareated}
\end{equation}

\section{Partition function evaluation}
\label{app:Zeval}

All the ingredients are now in place to evaluate the partition function, as this latter form of Equation \eqref{eq:Hsepareated} can be exponentiated as

\begin{equation}
    e^{-H(A)} = e^{\braket{A|\epsilon}+\sum_{k=1}^{r_+} \frac{\lambda_k}{2V} \braket{A|\omega_k}^2 + \sum_{k=r_+ +1}^r \frac{\lambda_k}{2V} \braket{A|\omega_k}^2} = e^{\braket{A|\epsilon}}\prod_{k=1}^{r_+} e^{\frac{\lambda_k}{2V} \braket{A|\omega_k}^2} \prod_{k=r_+ + 1}^{r} e^{\frac{\lambda_k}{2V} \braket{A|\omega_k}^2} 
\end{equation}

and this form is very amenable to an Hubbard Stratonovich transformation. This integral representation allows us to disentangle the squares as

\begin{equation}
    e^{a^2} = \frac{1}{\sqrt{2\pi}} \int_{-\infty}^{\infty} \,dx e^{-\frac{x^2}{2} + \sqrt{2}ax}
\end{equation}

if $a>0$ and

\begin{equation}
    e^{-a^2} = \frac{1}{\sqrt{2\pi}} \int_{-\infty}^{\infty} \,dx e^{-\frac{x^2}{2} + i\sqrt{2}ax}
\end{equation}

if $a<0$. This allows to write

\begin{equation}
    e^{-H(A)} = e^{\braket{A|\epsilon}}\prod_{k=1}^{r_+}\left[\frac{1}{\sqrt{2\pi}}\int \,dx e^{-\frac{x^2}{2} + x\sqrt{\frac{\lambda_k}{V}} \braket{A|\omega_k}}\right]\prod_{k=r_+ + 1}^{r}\left[\frac{1}{\sqrt{2\pi}}
\int \,dx e^{-\frac{x^2}{2} + i x \sqrt{\frac{\lambda_k}{V}} \braket{A|\omega_k}}\right]
\end{equation}

which we can write more compactly by retaining the local field $\epsilon_a$ outside the mode product. Using the convention
\begin{equation}
    \sqrt{\lambda_k} = 
    \begin{cases}
        |\sqrt{\lambda_k}| \text{ if } \lambda_k > 0,\\
        i |\sqrt{|\lambda_k|}| \text{ if } \lambda_k < 0,
    \end{cases}
\end{equation}
the Hubbard--Stratonovich representation becomes
\begin{equation}
    \begin{aligned}
    e^{-H(A)}
    &=
    \left(\frac{1}{\sqrt{2\pi}}\right)^{r}
    \int \cdots \int dx_1\cdots dx_r\,
    e^{-\sum_{k=1}^r x_k^2/2}
    e^{\braket{A|\epsilon}}
    \prod_{k=1}^{r}
    e^{x_k\sqrt{\frac{\lambda_k}{V}}\braket{A|\omega_k}} .
    \end{aligned}
    \label{eq:HScompact}
\end{equation}
Writing the scalar products explicitly, Eq.~\eqref{eq:HScompact} is
\begin{equation}
    e^{-H(A)}
    =
    \left(\frac{1}{\sqrt{2\pi}}\right)^{r}
    \int \cdots \int dx_1\cdots dx_r\,
    e^{-\sum_{k=1}^r x_k^2/2}
    \prod_{a=1}^{V}
    e^{A_a\left[
    \epsilon_a+
    \sum_{k=1}^{r}
    x_k\sqrt{\frac{\lambda_k}{V}}(\omega_k)_a
    \right]} .
    \label{eq:HSlink}
\end{equation}

Now we sum over the configurations. We therefore obtain
\begin{multline}
    Z = \sum_{\{A_a=0,1\}} e^{-H(A)} = \\
    = \left(\frac{1}{\sqrt{2\pi}}\right)^{r}
    \int \cdots \int dx_1\cdots dx_r\,
    e^{-\sum_{k=1}^r x_k^2/2}
    \sum_{\{A_a=0,1\}}
    \prod_{a=1}^{V}
    e^{A_a\left[
    \epsilon_a+
    \sum_{k=1}^{r}
    x_k\sqrt{\frac{\lambda_k}{V}}(\omega_k)_a
    \right]} = \\
    = \left(\frac{1}{\sqrt{2\pi}}\right)^{r}
    \int \cdots \int dx_1\cdots dx_r\,
    e^{-\sum_{k=1}^r x_k^2/2}
    \prod_{a=1}^{V}
    \left[
    1+
    e^{\epsilon_a+
    \sum_{k=1}^{r}
    x_k\sqrt{\frac{\lambda_k}{V}}(\omega_k)_a}
    \right] .
    \label{eq:Zexplicit}
\end{multline}

\section{Saddle Point evaluation of the Partition function}

Formula \eqref{eq:Zexplicit} is a $r$-dimensional integral in which all the modes are coupled through the product over the links, and it does not factorize into $r$ independent one-dimensional integrals. However, we can compute the large $V$ limit of the integral through a saddle point approximation, to show that the first order approximation of $Z$ becomes effectively separable, and that an effective single-link probability can be written as a function of the phase parameter $\mathbf{x^*}$.  

To make the saddle structure explicit, we introduce the effective action

\begin{equation} \Phi(\mathbf x)
=-\frac12\sum_{k=1}^{r}x_k^2 +\sum_{a=1}^{V}
\ln\left[
1+e^{
\epsilon_a+\sum_{k=1}^{r}
x_k\sqrt{\frac{\lambda_k}{V}}(\omega_k)_a
}\right].\label{eq:action}
\end{equation}

The partition function can then be written in the compact form

\begin{equation}
    Z=\left(\frac{1}{\sqrt{2\pi}}\right)^{r}
\int_{\mathbb R^r}d^r x\,e^{\Phi(\mathbf x)}
\end{equation}
which is in a form lending itself to a saddle point approximation. We first redefine the effective local fields as

\begin{equation}
    \epsilon'_a(\mathbf x)=\epsilon_a+
\sum_{k=1}^{r}x_k
\sqrt{\frac{\lambda_k}{V}}(\omega_k)_a
\end{equation}
and from the thermodynamical relation $p_a = \frac{\partial \ln Z}{\partial \epsilon_a}$ we can compute the corresponding effective link probabilities

\begin{equation}
    p_a(\mathbf x)=
\frac{e^{\epsilon'_a(\mathbf x)}}{1+e^{\epsilon'_a(\mathbf x)}}.
\end{equation}

With those auxiliary definition in place, we can start computing the partition function in the Large $V$ limit. Let us expand the action at the second order on the maximum of $\Phi$

\begin{equation}
    \Phi(\mathbf x)\simeq
\Phi(\mathbf x^*)+
\frac12(\mathbf x-\mathbf x^*)^{\mathsf T}
\mathbf{H}_{\Phi}(\mathbf x^*)(\mathbf x-\mathbf x^*),
\end{equation}

where $\mathbf{H}_{\Phi}(\mathbf{x^*})$ denotes the Hessian matrix of the action, whose elements are

\begin{equation}
    \left(\mathbf{H}_{\Phi}(\mathbf x^*)\right)_{kl}=\frac{\partial^2 \Phi}{\partial x_k \partial x_l}|_{\mathbf x^*} =-\delta_{kl}
+\sum_{a=1}^{V}
\frac{\lambda_k}{V}(\omega_k)_a(\omega_l)_a
p_a(\mathbf x^*)[1-p_a(\mathbf x^*)].
\end{equation}

Then, the partition function is approximated by

\begin{equation}
    Z \simeq \sum_{x^*}\left(\frac{1}{\sqrt{2\pi}}\right)^{r} e^{\Phi(\mathbf x^*)}
\int_{\mathbb R^r}d^r x\, e^{\frac12(\mathbf x-\mathbf x^*)^{\mathsf T}
\mathbf{H}_{\Phi}(\mathbf x^*)(\mathbf x-\mathbf x^*)}\label{eq:saddlezeta}.
\end{equation}

This is true for all the maxima (local and global) of $\Phi$, which entail the two conditions of $\mathbf{x^*}$ being a stationary point and $\mathbf{H}_{\Phi}(\mathbf{x^*})$ being definite negative. The first condition, \textit{i.e.} the fact that the gradient of $\Phi$ should be null in $\mathbf{x^*}$, can be recast in the entrywise equation

\begin{equation}
    x_k^*=
\sum_{a=1}^{V}
\sqrt{\frac{\lambda_k}{V}}(\omega_k)_a
p_a(\mathbf x^*) \quad \forall k \label{eq:stationarity}
\end{equation}
These coupled self-consistency equations define the critical points \(\mathbf x^*\) of the  action $\Phi$. The solutions may be unique or multiple, depending on the interaction spectrum and the local fields \(\epsilon_a\). If multiple solutions are in place, the contribution coming from each one of them has to be retained, effectively splitting the partition function into a sum of more contributions each coming from a different value of $x^*$, which then acts as a phase indicator parameter. However, when two or more locally stable maxima exist, their relative contributions to \(Z\) determine which phases are thermodynamically relevant, and which are suppressed in the large $V$ limit. Coexistence between two saddles \(\mathbf x^{*(i)}\) and \(\mathbf x^{*(j)}\) requires

\begin{equation}
    \Phi(\mathbf x^{*(i)})
\sim\Phi(\mathbf x^{*(j)}),
\end{equation}

because otherwise the contribution coming from one local maximum is exponentially suppressed in the large $V$ limit with respect to the global one.

Finally, if the integral on the right of \eqref{eq:saddlezeta} can be evaluated by diagonalizing the Hessian, yielding $r$ decoupled Gaussian integrals. By integrating those, one gets 

\begin{equation}
    Z \simeq \sum_{x^*} \frac{e^{\Phi(\mathbf x^*)}}{\sqrt{\det \mathbf{H}_{\Phi}(\mathbf x^*)}}\label{eq:zIIorder}.
\end{equation}

\subsection{Effective separability of the partition function}
\label{app:zsep}

A separable partition function in the form

\begin{equation}
    Z = \prod_{a=1}^V z_a \label{eq:sep}
\end{equation}

is what one would like to work with, since it has a straightforward connection to the probability of link formation. As a matter of facts, when the link variables are not coupled there is an identification between the terms in the product of $Z$ and the bare link probabilities. This can be shown from 

\begin{equation}
    Z = \sum_{\{A_{a}=0,1\}} e^{-H(A)} = \sum_{\{A_{a}=0,1\}} \prod_{a=1}^Ve^{ h^{(1)}_{a}A_{a}} = \prod_{a=1}^V \left(1 + e^{h^{(1)}_a}\right) = \prod_{a=1}^V \frac{1}{1-p_a}
\end{equation}

The saddle-point representation of eq.\eqref{eq:saddlezeta} also provides a direct connection between the interacting network model and a family of first-order Random Network Models. If one retains only the first order term in the approximation, it is easy to see that the Partition function is straightforwardly separable on each saddle point

\begin{equation}
    Z \simeq \left(\frac{1}{\sqrt{2\pi}}\right)^{r}\sum_{x^*} \left[e^{-\sum_{k=1}^{r} \frac{(x^*_k)^2}{2} } \prod_{a=1}^{V}\left(1 + e^{\epsilon_a'(\mathbf{x^*})}\right)\right]
\end{equation}
in which the single link contribution can be identified as

\begin{equation}
    z_a(\mathbf{x^*}) = \underbrace{\left(\frac{1}{\sqrt{2\pi}}\right)^{r/V} e^{-\sum_{k=1}^{r} \frac{(x^*_k)^2}{2V}}}_{c(\mathbf{x^*})}\left[1 + e^{\epsilon_a + \sum_{k=1}^{r}
x_k^*\sqrt{\frac{\lambda_k}{V}}(\omega_k)_a}\right]
\end{equation}
This is the proof of the claim made in the main text, that is the approximate partition function becomes separable conditionally on the value of $\mathbf{x^*}$, which identifies the phase the model is in.  

\subsection{Goodness of approximation}
\label{sec:goodapp}
The validity of the first-order saddle-point approximation in the large $V$ limit can be assessed by examining the scaling of the Gaussian correction in Eq.~\eqref{eq:zIIorder}. For a non-degenerate maximum $\mathbf{x}^*$, let $h_1,\ldots,h_r$ denote the eigenvalues of the Hessian. The contribution of this saddle to the logarithm of the partition function is, up to the normalization already included in Eq.~\eqref{eq:zIIorder},
\begin{equation}
    \log Z_{\mathbf{x}^*}
    \simeq
    \Phi(\mathbf{x}^*)
    -\frac{1}{2}\sum_{k=1}^{r}\log\mu_k .
\end{equation}
Assuming that the Hessian eigenvalues remain uniformly bounded
away from zero and infinity as $N\to\infty$, the fluctuation
correction is $O(r)$. Since the number of links scales as
$V=N(N-1)/2=O(N^2)$, this correction is subextensive both when
$r$ is finite, in which case it is $O(1)$, and when $r=O(N)$,
in which case it is at most $O(N)$. Therefore, the Gaussian contribution does not modify the leading thermodynamic free-energy density. Under these conditions, the
first-order saddle-point approximation is therefore asymptotically
exact at the level of $V^{-1}\log Z$ for both finite and
linearly growing $r$. 

\subsection{Zero as maximizer for sparse graphs}

From the stationarity condition \eqref{eq:stationarity} we can proceed to evaluate the condition for which the origin $\mathbf{x}=0$ is a stationary point. Indeed, by evaluating \eqref{eq:stationarity} in zero we get

\begin{equation}
    0 =  \sqrt{\frac{\lambda_k}{V}}\sum_{a=1}^V (\omega_k)a \frac{e^{\epsilon_a}}{1+e^{\epsilon_a}} \label{eq:zeromax}
\end{equation}

The origin $\mathbf{x}=0$ is then a maximal if the above condition holds. There are several conditions that make this zero, one is the regular case discussed in the main text, another example is when the graph is sparse, \textit{i.e.} $\epsilon_a \to -\infty$, which is incidentally a common feature of real-world networks.

\section{A basis for the community structure induced coupling}
\label{app:SBMbasis}

We can write a rank three matrix in the following form

\begin{equation}
    \beta = \lambda_1 \ket{\omega_1}\bra{\omega_1} + \lambda_2 \ket{\omega_2}\bra{\omega_2} + \lambda_3 \ket{\omega_3}\bra{\omega_3}
\end{equation}

where the eigenvectors are

\begin{equation}
    \bra{\omega_1}= \frac{1}{\sqrt{n_1}}(\underbrace{\left(\omega_1\right)_1, \ldots, \left(\omega_1\right)_1}_{V_1},\underbrace{\left(\omega_1\right)_2 , \ldots, \left(\omega_1\right)_2 }_{V_2},\underbrace{\left(\omega_1\right)_3 , \ldots, \left(\omega_1\right)_3 }_{V_3})
\end{equation}

\begin{equation}
    \bra{\omega_2}= \frac{1}{\sqrt{n_2}}(\underbrace{\left(\omega_2\right)_1, \ldots, \left(\omega_2\right)_1}_{V_1},\underbrace{\left(\omega_2\right)_2 , \ldots, \left(\omega_2\right)_2 }_{V_2},\underbrace{\left(\omega_2\right)_3 , \ldots, \left(\omega_2\right)_3 }_{V_3})
\end{equation}

\begin{equation}
    \bra{\omega_3}= \frac{1}{\sqrt{n_3}}(\underbrace{\left(\omega_3\right)_1, \ldots, \left(\omega_3\right)_1}_{V_1},\underbrace{\left(\omega_3\right)_2 , \ldots, \left(\omega_3\right)_2 }_{V_2},\underbrace{\left(\omega_3\right)_3 , \ldots, \left(\omega_3\right)_3 }_{V_3}).
\end{equation}

where the norms are

\begin{equation}
    n_1 = V_1 \left(\omega_1\right)_1^2 + V_2 \left(\omega_1\right)_2^2 + V_3 \left(\omega_1\right)_3^2
\end{equation}

\begin{equation}
    n_2 = V_1 \left(\omega_2\right)_1^2 + V_2 \left(\omega_2\right)_2^2 + V_3 \left(\omega_2\right)_3^2
\end{equation}

\begin{equation}
    n_3 = V_1 \left(\omega_3\right)_1^2 + V_2 \left(\omega_3\right)_2^2 + V_3 \left(\omega_3\right)_3^2
\end{equation}

The eigenvalues can be computed from the secular equation

\begin{equation}
    \beta\ket{\omega_i} = \lambda_i \ket{\omega_i}
\end{equation}

so that

\begin{equation}
    \begin{cases}
        \lambda_1 = V_1 \beta_1 + V_2 \beta_{12}\frac{\left(\omega_1\right)_2}{\left(\omega_1\right)_1} + V_3 \beta_{13}\frac{\left(\omega_1\right)_3}{\left(\omega_1\right)_3}\\
        \lambda_2 = V_1 \beta_1 + V_2 \beta_{12}\frac{\left(\omega_2\right)_2}{\left(\omega_2\right)_1} + V_3 \beta_{13}\frac{\left(\omega_2\right)_3}{\left(\omega_2\right)_1}\\
        \lambda_3 = V_1 \beta_1 + V_2 \beta_{12}\frac{\left(\omega_3\right)_2}{\left(\omega_3\right)_1} + V_3 \beta_{13}\frac{\left(\omega_3\right)_3}{\left(\omega_3\right)_1}
    \end{cases}
\end{equation}

The equation for the spectral decomposition defines the relations between the eigenvector components and the $\beta$ parameters

\begin{equation}
    \begin{cases}
        \beta_1 = \lambda_1 \left(\omega_1\right)_1^2 + \lambda_2 \left(\omega_2\right)_1^2 + \lambda_3 \left(\omega_3\right)_1^2 \\
        \beta_2 = \lambda_1 \left(\omega_1\right)_2^2 + \lambda_2 \left(\omega_2\right)_2^2 + \lambda_3 \left(\omega_3\right)_2^2 \\
        \beta_3 = \lambda_1 \left(\omega_1\right)_3^2 + \lambda_2 \left(\omega_2\right)_3^2 + \lambda_3 \left(\omega_3\right)_3^2 \\
        \beta_{12} = \lambda_1 \left(\omega_1\right)_1 \left(\omega_1\right)_2 + \lambda_2 \left(\omega_2\right)_1 \left(\omega_2\right)_2 + \lambda_3 \left(\omega_3\right)_1 \left(\omega_3\right)_2 \\
        \beta_{13} = \lambda_1 \left(\omega_1\right)_1 \left(\omega_1\right)_3 + \lambda_2 \left(\omega_2\right)_1 \left(\omega_2\right)_3 + \lambda_3 \left(\omega_3\right)_1 \left(\omega_3\right)_3 \\
        \beta_{23} = \lambda_1 \left(\omega_1\right)_2 \left(\omega_1\right)_3 + \lambda_2 \left(\omega_2\right)_2 \left(\omega_2\right)_3 + \lambda_3 \left(\omega_3\right)_2 \left(\omega_3\right)_3
    \end{cases}
\end{equation}

and these are supplemented by the orthogonality equations so that

\begin{equation}
    \begin{cases}
        V_1 \left(\omega_1\right)_1 \left(\omega_2\right)_1 + V_2 \left(\omega_1\right)_2 \left(\omega_2\right)_2 + V_3 \left(\omega_1\right)_3 \left(\omega_2\right)_3 = 0 \\
        V_1 \left(\omega_1\right)_1 \left(\omega_3\right)_1 + V_2 \left(\omega_1\right)_2 \left(\omega_3\right)_2 + V_3 \left(\omega_1\right)_3 \left(\omega_3\right)_3 = 0 \\
        V_1 \left(\omega_2\right)_1 \left(\omega_3\right)_1 + V_2 \left(\omega_2\right)_2 \left(\omega_3\right)_2 + V_3 \left(\omega_2\right)_3 \left(\omega_3\right)_3 = 0
    \end{cases}
\end{equation}

which completely defines the $\omega$ and the eigenvalues. We also define the quantities $\alpha = \frac{V_1}{V}$, $\beta = \frac{V_2}{V}$ and $\gamma = \frac{V_3}{V}$. 

The action functional in this case becomes
\begin{multline}
    \Phi(x,y,z) = -\frac{x^2 + y^2 + z^2}{2} + V\left[\alpha \ln \left(1 + e^{\epsilon_1 + x\sqrt{\frac{\lambda_1}{V n_1}} \left(\omega_1\right)_1 + y \sqrt{\frac{\lambda_2}{V n_2}}\left(\omega_2\right)_{1} + z \sqrt{\frac{\lambda_3}{V n_3}}\left(\omega_3\right)_{1}}\right)\right. \\
    + \left.\beta \ln \left( 1 + e^{\epsilon_2 + x\sqrt{\frac{\lambda_1}{V n_1}} \left(\omega_1\right)_2 + y \sqrt{\frac{\lambda_2}{V n_2}}\left(\omega_2\right)_{2} + z \sqrt{\frac{\lambda_3}{V n_3}}\left(\omega_3\right)_{2}}\right) + \gamma \ln \left( 1 + e^{\epsilon_3 + x\sqrt{\frac{\lambda_1}{V n_1}} \left(\omega_1\right)_3 + y \sqrt{\frac{\lambda_2}{V n_2}}\left(\omega_2\right)_{3} + z \sqrt{\frac{\lambda_3}{V n_3}}\left(\omega_3\right)_{3}} \right) \right]
\end{multline}

and the partition function to be computed is again, with a saddle point approximation,

\begin{equation}
    Z = \left(\frac{1}{2 \pi}\right)^{3/2} \int \,dx dy dz \, e^{\Phi(x,y,z)} \simeq \sum_{x^*} \frac{e^{\Phi(\mathbf x^*)}}{\sqrt{\det \mathbf{H}_{\Phi}(\mathbf x^*)}}
\end{equation} 

To identify the different maxima where the phases might emerge, one needs again to look at the critical points of the action where its gradient vanishes, as in eq \eqref{eq:stationarity}. This was done numerically, and we show in the main text and in figure \ref{fig:morephases} a subset of the possible phase structure that emerge from the analysis of the critical point of $\Phi$.

\end{document}